\documentclass[journal]{IEEEtran}

\usepackage{xcolor,soul,framed} %,caption

\colorlet{shadecolor}{yellow}
\usepackage[pdftex]{graphicx}
\graphicspath{{../pdf/}{../jpeg/}}
\DeclareGraphicsExtensions{.pdf,.jpeg,.png}

\usepackage[cmex10]{amsmath}
\usepackage{array}
\usepackage{mdwmath}
\usepackage{mdwtab}
\usepackage{eqparbox}
\usepackage{url}
\usepackage{cite}

\usepackage{ragged2e} 
\usepackage{booktabs,makecell, multirow, tabularx}

\usepackage[ruled,linesnumbered]{algorithm2e}
\usepackage[numbers]{natbib}

\begin{document}
\bstctlcite{IEEEexample:BSTcontrol}
    \title{Cooperative Driving of Connected Autonomous Vehicles in Heterogeneous Mixed Traffic: A Game Theoretic Approach}
  \author{Shiyu Fang,
      Peng Hang,~\IEEEmembership{Member,~IEEE,} Chongfeng Wei,~\IEEEmembership{Member,~IEEE,}\\
      Yang Xing,~\IEEEmembership{Member,~IEEE,}
      and Jian Sun% <-this % stops a space
\thanks{This work was supported in part by the Natural Science Foundation of China (52232015 and 52125208), the Fundamental Research Funds for the Central Universities (No. 2022-5-ZD-02), and the Zhejiang Lab (2021NL0AB02).}
\thanks{S. Fang, P. Hang and J. Sun are with the Department of
Traffic Engineering and Key Laboratory of Road and Traffic Engineering,
Ministry of Education, Tongji University, Shanghai 201804, China. (e-mail: \{2111219, hangpeng, sunjian\}@tongji.edu.cn)}
\thanks{C. Wei is with the School of Mechanical and Aerospace Engineering at Queen’s University Belfast, BT7 1NN Belfast, UK. (email: c.wei@qub.ac.uk)}
\thanks{Y. Xing is with the School of Aerospace, Transport, and Manufacturing, Cranfield University, UK, MK43 0AL. (e-mail: yang.x@cranfield.ac.uk)}
\thanks{Corresponding author: P. Hang}
\thanks{This work has been submitted to the IEEE for possible pubilcation. Copyright may transferred without notice, after which this version may no longer be accessible}
}

\markboth{}{Roberg \MakeLowercase{\textit{et al.}}: High-Efficiency Diode and Transistor Rectifiers}

% ====================================================================
\maketitle

% === ABSTRACT ====================================================================
% =================================================================================
\begin{abstract}
%\boldmath
High-density, unsignalized intersection has always been a bottleneck of efficiency and safety. The emergence of Connected Autonomous Vehicles (CAVs) results in a mixed traffic condition, further increasing the complexity of the transportation system. Against this background, this paper aims to study the intricate and heterogeneous interaction of vehicles and conflict resolution at the high-density, mixed, unsignalized intersection. Theoretical insights about the interaction between CAVs and Human-driven Vehicles (HVs) and the cooperation of CAVs are synthesized, based on which a novel cooperative decision-making framework in heterogeneous mixed traffic is proposed. Normalized Cooperative game is concatenated with Level-k game (NCL game) to generate a system optimal solution. Then Lattice planner generates the optimal and collision-free trajectories for CAVs. To reproduce HVs in mixed traffic, interactions from naturalistic human driving data are extracted as prior knowledge. Non-cooperative game and Inverse Reinforcement Learning (IRL) are integrated to mimic the decision making of heterogeneous HVs. Finally, three cases are conducted to verify the performance of the proposed algorithm, including the comparative analysis with different methods, the case study under different Rates of Penetration (ROP) and the interaction analysis with heterogeneous HVs. It is found that the proposed cooperative decision-making framework is beneficial to the driving conflict resolution and the traffic efficiency improvement of the mixed unsignalized intersection. Besides, due to the consideration of driving heterogeneity, better human-machine interaction and cooperation can be realized in this paper. 

\end{abstract}

% === KEYWORDS ====================================================================
% =================================================================================

% === KEYWORDS ====================================================================
% =================================================================================

% \fi
\IEEEpeerreviewmaketitle

% ====================================================================
% ====================================================================
% ====================================================================

% === I. INTRODUCTION =============================================================
% =================================================================================
\section{Introduction}

\IEEEPARstart{T}{he} past decade has seen the rapid development of Connected Autonomous Vehicles (CAVs). CAV has a wider perception range and smaller perception error which attach it the potential to perceive danger earlier and even be able to actively cooperate with other vehicles to resolve conflict by properly arranging Right of Way (ROW). However, there still exist some challenges. On account of the complexity of real-world situations, CAVs may behave unreasonably when interacting with Human Vehicles (HVs) which may lead to collision or deadlock. There is a further question about how to ascend system performance by coordinating the relationship between vehicles. Not to mention that CAV control in mixed traffic containing CAVs and heterogeneous HVs is still a continuing cause for concern. 

For decades, one of the ideas to ameliorate the above problem is to explore the latent rules and patterns from human interactions. Numerous studies have analyzed drivers’ decision-making logic, cognitive structure, and inherent characteristics to design a human-like CAV. Game theory formulates the interaction between incentive structures and the relationship between players' strategies which has been widely used to replicate social decision making \cite{schwarting2019social, camerer2006does}. Except for the advantage of interpretability, game theory is also suitable for dissimilar scenarios such as roundabouts \cite{tian2018adaptive}, on-ramp merging areas \cite{xu2020nash, garzon2019game}, and unsignalized intersections \cite{tian2020game, wang2020performance}. Instead of treating surrounding vehicles as moving obstacles, the game theoretic approach was adapted to mimic human behavior by dynamically interacting with surrounding vehicles for automatic lane changing \cite{yu2018human}. Level-k game was combined with vehicle controller \cite{li2018game} and Monte Carlo tree search (MTCS) \cite{karimi2020receding} respectively to model the time-extended, multi-step, and interactive decision making of CAV. Meanwhile, identifying real-time intentions through human cognitive structure plays an important role in developing human-centric intelligent vehicles. Therefore, \citet{chu2022understanding} explored the mechanisms behind distracted driving intentions based on Stimulus-Organism-Response (SOR) cognitive theory. In addition, Artificial Neural Network (ANN) was designed by mimicking the mechanism by which information is transmitted in the animal neural network and it was used to predict drivers' intentions in on-ramp merging \cite{el2021novel}, lane changing \cite{yu2020deep}, and left-turning \cite{yao2021deep}. \citet{bi2012queuing} proposed a queuing network cognitive architecture to infer intentions with and without distraction tasks and the experiments carried out on a driving simulator have shown a good performance in inferring typical and rapid lane-changing behavior intentions. Same to the diverse and fickle intentions, drivers themselves also possess different characteristics. \citet{huang2021driving} adopted Inverse Reinforcement Learning (IRL) to model the different driving behavior from naturalistic human driving data. \citet{schwarting2019social} addressed Social Value Orientation (SVO), which quantifies the degree of an agent's selfishness or altruism, and then modeled the interactions between agents as a best-response game. Despite a number of works that have shown that the interaction between CAVs and HVs can be solved by a best-response game, a rational collective decision might be the opposite of the socially optimum \cite{chremos2020game}. Therefore, a solution to tackle this phenomenon is to establish an institutional arrangement that can optimize the system performance via cooperative driving.

The cooperative driving of CAVs has been widely studied. Cooperation refers to the decision-making process of vehicles aimed at orchestrating vehicles' actions so as to achieve a goal that can not be achieved by each vehicle in isolation \cite{mariani2021coordination}. Furthermore, CAVs cooperative approaches can be classified into centralized, negotiation, agreement, and emergent according to CAVs' extent of autonomy. In the centralized control, a coordinator will act as an intersection manager to reserve certain space-time for each approaching vehicle \cite{yu2019managing} or to decide all vehicles' strategies for at least one global task \cite{rios2016survey}. However, the capability of the coordinator is a potential bottleneck. \citet{zhou2022reasoning} proposed a situation-aware Reasoning Graph (RG) and combined it with some rules for maneuver compatibility and social interaction customs to quickly search for speed profiles that follow the reasoned situations. Although \citet{xu2019cooperative} combined MCTS with heuristic rules to find a nearly global-optimal passing order for CAVs and results showed that it can keep a good trade-off between system performance and computation flexibility. Under centralized control, CAVs only passively follow the instruction of the coordinator so their potentials are far from being tapped. Negotiation and agreement approaches use fixed and dynamics protocols respectively to allow CAVs to communicate with others. \citet{carlino2013auction} proposed a decentralized auction-based autonomous intersection management scheme to permit vehicle passage based on their value of time. \citet{vu2018decentralised} further migrated auctions to a cellular automaton model. However, these cooperation approaches are infeasible for HVs because the action of human is uncontrollable. Simultaneously, during communication, CAVs adhering to a meticulously designed protocol will further cause robustness problems. Instead of directly communicating with other vehicles or following any predefined cooperation protocol, in emergent approaches, each vehicle makes its decision by estimating the rationality of others and deducing their actions based on the current state. \citet{wang2020performance} established a cooperation model for agents with different priorities through game theory. In addition, previous work also hybridized objective-novelty evolutionary search for synthesizing CAV cooperative behavior on CAVs-only roads \cite{huang2020evolutionary}. Experiments showed that desired cooperative driving behavior emerged when multiple vehicles interacted, but their results failed to generalize to new roads. To sum up, the above research focused on CAVs-only environment cooperation. Nevertheless, real-world traffic flow will be mixed with HVs for at least 40 years according to \cite{litman2017autonomous}, so the effectiveness under a mixed traffic environment has yet to be proven. 

Aiming at either the interactions between CAV and HV or the cooperation methods in a CAVs-only environment merely leads to a half-baked CAV, their performance under complex interactive scenarios should be further evaluated. So far, there has been little discussion about the verification and validation of CAV in mixed traffic. \citet{wang2015game} demonstrated his controller by adding background traffic flow but only four vehicles at most were involved in their two-lane merging scenario. Besides, human-in-the-loop trail on the driving simulator is also a practical way to replicate mixed traffic. \citet{sadigh2018planning} pioneering excavated the ability for CAV to actively gather the information of surrounding HV, and experiments were carried out on a driving simulator. The fly in the ointment is that experiments only consider one-on-one interaction. Furthermore, multiple human-machine interactions seem unpractical on the driving simulator because of the equipment limitations. Therefore, existing studies about mixed traffic are mainly low-density which is an extraordinary simplification. While real-world heterogeneous mixed traffic is far more hazardous because drivers' preferences will drive them to make different decisions.

To address the aforementioned challenges, the cooperative driving of CAVs in heterogeneous mixed traffic conditions is studied. The contributions of this paper are presented as follows:

1. This paper proposes a novel cooperative driving framework that enables CAVs to cooperate in high-density, mixed, unsignalized intersections and coordinate with heterogeneous human drivers, in favor of driving conflict resolution, human-machine interaction augmentation and traffic efficiency improvement.

2. The cooperative decision making of CAVs is realized by concatenating Normalized Cooperative game and Level-k game (NCL game), which outputs the k-allocation that corresponds to maximum system overall efficiency. Then Lattice planner is combined to generate an optimal and collision-free CAV trajectory. The cooperation performance of CAVs under different Rates of Penetration (ROP) is verified.

3. To simulate the human-machine interaction in authentic mixed traffic, heterogeneous HV decisions are modelled. By clustering, driver classification and composition are obtained and serve as prior knowledge. Non-cooperative game is used to reproduce driver's rational decision under an interactive environment and IRL is utilized to calibrate different drivers' decision-making preferences. Finally, heterogeneous HVs are introduced to the simulation experiment to examine the effectiveness and robustness of the cooperative driving framework.

The rest of the paper is organized as follows. Section \uppercase\expandafter{\romannumeral2} introduces the main problem of our research and the corresponding cooperative driving framework to solve it. Reproducing heterogeneous HV decision is described in Section \uppercase\expandafter{\romannumeral3}. Section \uppercase\expandafter{\romannumeral4} discusses the process of CAV cooperative decision making and trajectory planning. In Section \uppercase\expandafter{\romannumeral5}, we validate our framework through several simulation experiments. Finally, conclusions are made in Section \uppercase\expandafter{\romannumeral6}.

% === II. Problem statement and algorithm framework ========================
% =================================================================================
\section{Problem Statement and Cooperative Driving Framework}
CAVs are still much-maligned due to their intricate maneuvers and deficient control methods. Deadlock and collision may happen when CAVs interact with others in complex, interactive scenarios. Among all urban traffic scenarios, unsignalized intersections are the most challenging owing to their numerous conflict points and fickle interaction opponents. Furthermore, human drivers are heterogeneous so they may adopt various maneuvers in the same circumstances based on their decision-making preferences. Therefore, CAVs always tend to passive defensive action such as excessive deceleration which leads to inefficiencies, or even worse, to the loss of trust in CAVs.

% =======
% FIG. 01
% =======
\begin{figure}
  \begin{center}
  \centerline{\includegraphics[width=3.5in]{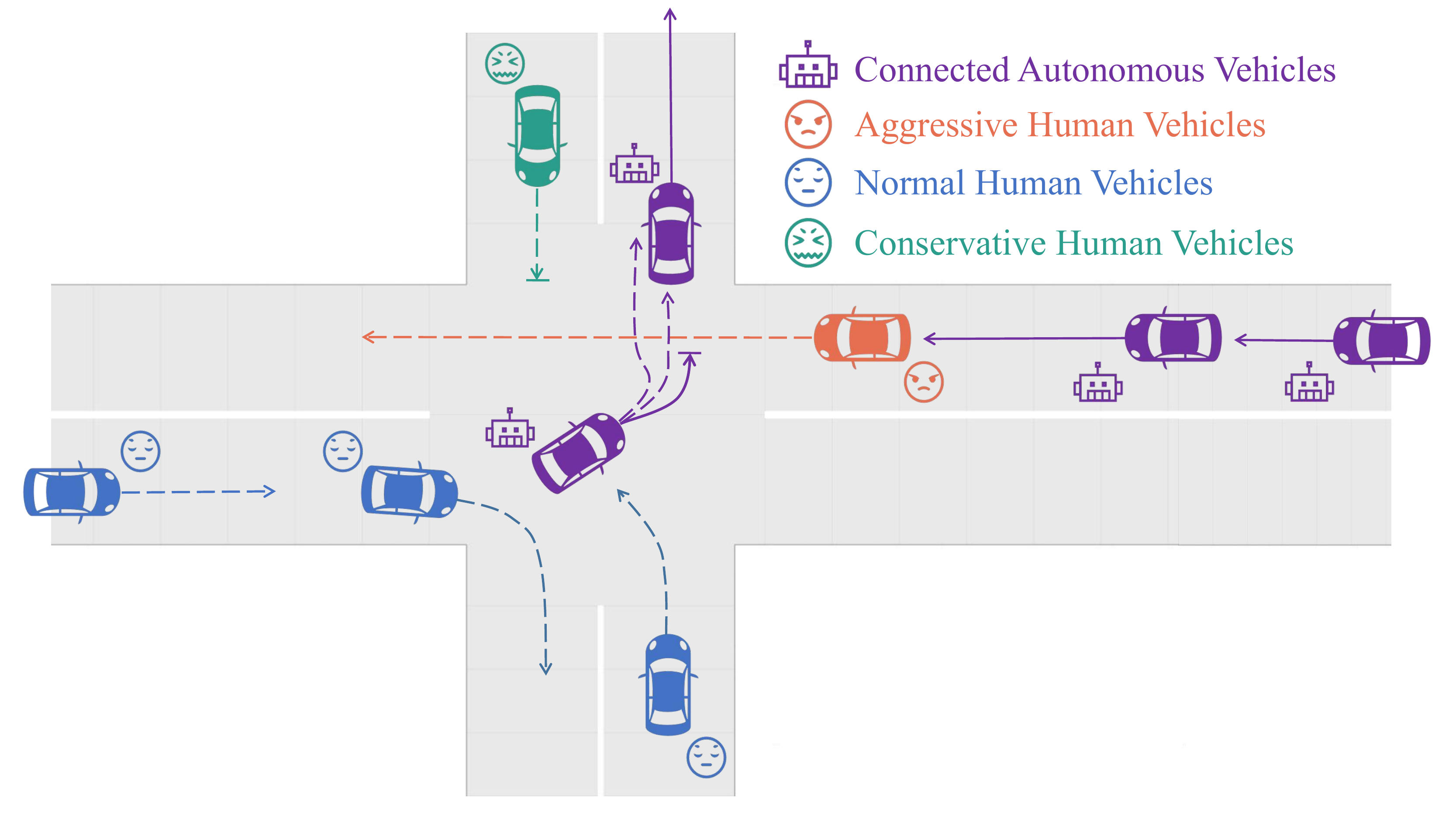}}
  \caption{Thumbnail of the high-density, unsignalized intersection with heterogeneous HVs involved.}\label{problem}
  \end{center}
\end{figure}

However, few studies have focused on multiple human-machine interactions which will be ubiquitous in the coming future. Hence the effectiveness and robustness of the algorithm for CAVs remain unclear. We hold the opinion that investigating the interactions of CAVs and HVs in sophisticated scenarios is of great significance to the validation and application of our cooperative driving algorithm for CAVs. Therefore, the main problem of this paper is to enable CAVs to properly respond to or even actively coordinate with heterogeneous HVs in a high-density, unsignalized intersection. Fig.~\ref{problem} shows a thumbnail of the intersection.

% =======
% FIG. 02
% =======
\begin{figure*}
  \begin{center}
  \centerline{\includegraphics[width=7in]{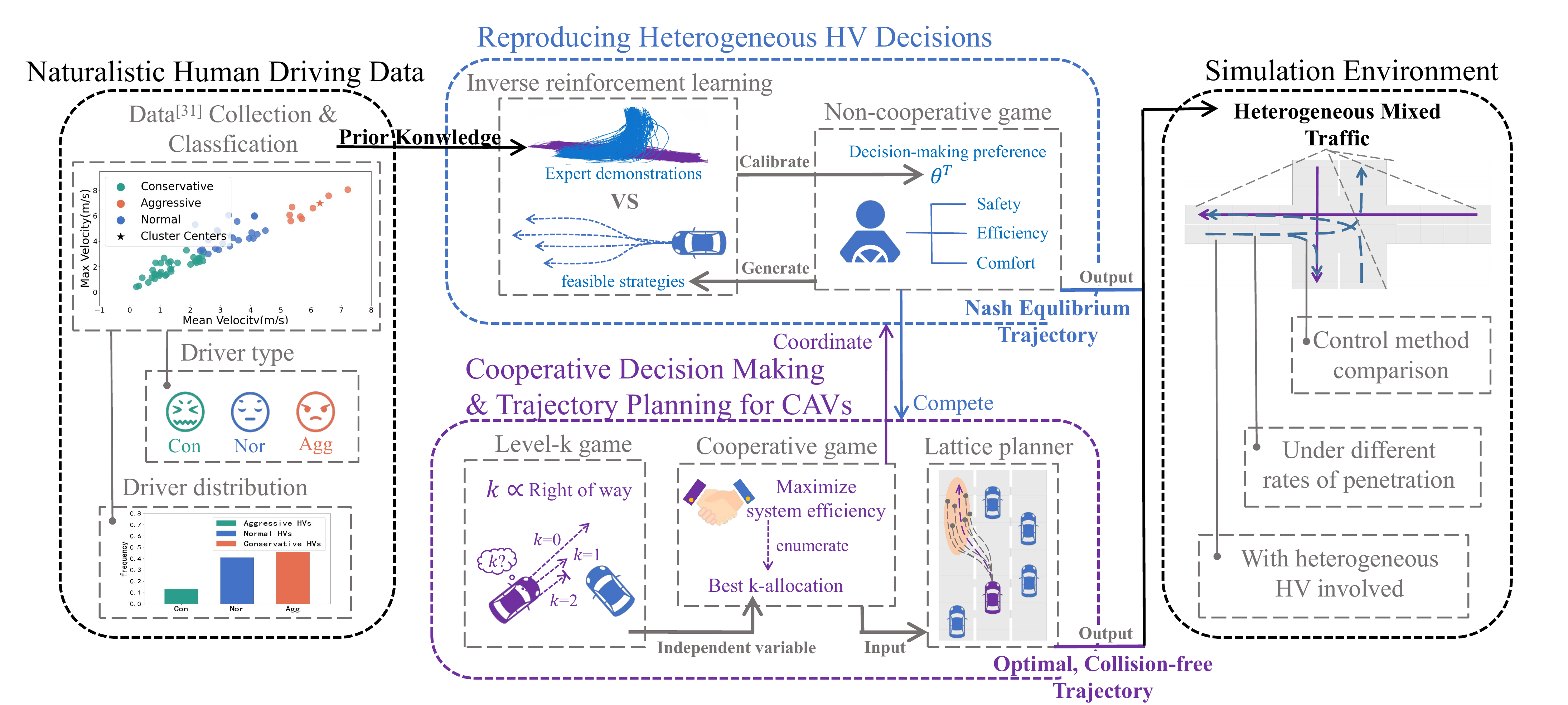}}
  \caption{Cooperative driving framework in mixed traffic.}\label{framework}
  \end{center} 
\end{figure*}

To achieve the above purpose, an elaborate framework is proposed, shown in Fig.~\ref{framework}. Firstly, naturalistic human driving data are collected and analyzed for investigating the driver decision type and its distribution from a real-world intersection. Besides, it has to be noted that modeling heterogeneous drivers mammoth project. Given that this paper is primarily concerned with establishing a cooperative driving framework for CAV, we turn to reproduce the heterogeneous decision for simplification. Then a series of feasible trajectories are generated through non-cooperative game and compare them with expert demonstrations from naturalistic human driving data. Through maximum entropy IRL, different decision-making preferences $\boldsymbol{\theta}^{T}$ are calibrated. We then reproduce the decision of heterogeneous HVs with Nash Equilibrium solution and generate the corresponding next state through vehicle dynamics. 

In addition, for the cooperative decision making and trajectory planning of CAVs, the three-layer hierarchy that most autonomous robot controls are using to generate the safe and continuous state is adopted \cite{vaskov2019towards, liu2022three}. At the top of the hierarchy, level-k game is introduced to imitate the human reasoning depths and also served as a next-layer's independent variable. Then, based on cooperative game, k-allocation which leads to system optimum can be achieved by enumerating. At the bottom layer, Lattice planner will generate an optimal and collision-free trajectory that conforms to the vehicle's dynamic constraints.

Mixed traffic in the simulation will be updated every $\Delta t=0.1s$ through vehicles dynamically competing and coordinating with each other. Finally, we validate our framework with experiments, including the comparative analysis with different methods, the case study under different Rates of Penetration (ROP) and the interaction analysis with heterogeneous HVs.

% === III. Reproducing heterogeneous HVs in mixed traffic===================
% =================================================================================
\section {Reproducing Heterogeneous HV Decisions in Mixed Traffic}

Despite the uncertainty due to the absence of traffic lights, unpredictable and uncontrollable HV latent intention of decision may lead to even greater strait for CAVs. In order to validate the cooperative driving algorithm of CAVs in heterogeneous mixed traffic. In this section, sophisticated real-world drivers' demonstrations were extracted for reproducing heterogeneous HV decisions. 

The main process consists of four steps. Firstly, interaction data from a real-world intersection were collected. Secondly, drivers' decisions were classified into three groups by K-means cluster. Then, non-cooperative game, known as a promising model to reproduce the process of human decision making, was selected to guide the generation of HVs' decision. Finally, the decision-making preference of each group was calibrated through IRL.

\subsection {Data Collection}
As mentioned before, investigating the interactions of HVs is of great use to establish either HVs or human-like CAVs. In our paper, data acquisition was carried out at an intersection in Shanghai, China: XianXia Rd-JianHe Rd (XXJH) based on previous research of \citet{ni2016evaluation}. Then interaction event was extracted if its Post Encroachment Time (PET) was less than 3s.

After screening, 79 of 131 interaction events were retained. Meanwhile, the trajectories of the HVs were extracted through a semi-automated video image processing tool and then served as prior knowledge. Furthermore, their speed and acceleration were calculated for clustering.

\subsection {Drivers' Decisions Classification by K-means Cluster}
As is known to all that drivers' decisions can be divided into several types such as aggressive, normal, and conservative. In order to reproduce drivers with different decision-making preferences. The first priority was to determine how many decision clusters the drivers at the intersection of XXJH can be divided into. 

Based on the result of the elbow method, a simple but versatile theory to determine the number of clusters using the Sum of Squared Errors (SSE), three clusters work best for our data. Considering that speed and acceleration are the most intuitive manifestation of decision making during driving, the average, maximum, minimum, and standard deviation of speed and acceleration of vehicles were taken as the clustering parameters based on \citet{chen2019driving}. K-means clustering was then used to classify naturalistic human driving data samples in XXJH.

% =======
% TABLE. 01
% =======
\begin{table}[htbp]
    \centering
    \caption{Clustering Centers of XXJH Data}
    \begin{tabular}{c|c c c c}
            \hline
    \multirow{2}{*}{Params} & \multicolumn{4}{c}{$V(m/s)$} \\
    \cline{2-5}
        & mean & max & max & min \\
    \hline
    Cluster-1 & 3.31 & 4.42 & 2.36 & 0.66 \\
    Cluster-2 & 1.34 & 1.60 & 0.99 & 0.27 \\
    Cluster-3 & 6.29 & 6.98 & 5.40 & 0.50 \\
    \hline
    \multirow{2}{*}{Params} & \multicolumn{4}{c}{$A(m/s^{2})$} \\
    \cline{2-5}
         & mean & max & max & min \\
    \hline
    Cluster-1 & -0.07 & 0.49 & -0.72 & 0.37 \\
    Cluster-2 & 0.08 & 0.55 & -0.34 & 0.28 \\
    Cluster-3 & -0.77 & -0.30 & -1.14 & 0.28 \\
    \hline
    \end{tabular}
    \label{table 1}
\end{table}

After the clustering algorithm converged, the center of each cluster can be obtained, as shown in Table.~\ref{table 1}. Since it was not the main concern of our study, we simply assumed that speed was the most important variable affecting the aggressiveness of a decision while driving based on \citet{huang2018effects}. Therefore, the drivers' decisions in the 79 interaction events were divided into three clusters: normal, conservative, and aggressive. Furthermore, their trajectories were used as expert demonstrations and then compared with HVs' feasible trajectories to calibrate different decision-making preferences. In this paper, non-cooperative game was chosen to estimate possible decisions human may adopt when confronted with other drivers because it achieves more realistic human behavior
when performing conflicting maneuvers at intersections \cite{rahmati2021helping, cheng2019vehicle}. Together with vehicle dynamics, feasible trajectories of each frame can be concluded.

\subsection {HVs' Feasible Trajectories Generation through Non-cooperative Game}
As mentioned before, game theory describes human as a rational decision-maker who takes action dependencies into account. Hence, there is a growing application in the field of modeling human decision. In this paper, we regard human drivers as rational decision-makers and are aware of the consequence that their actions will influence other drivers who have conflict with them in temporal and spatial dimensions. Non-cooperative game was therefore established to mimic how drivers conjecture and compete with each other while driving in sharing space.

We considered a sampling interval, $\Delta t=0.1s$, and an action set including six strategies that represent common driving maneuvers in urban traffic based on \citet{tian2018adaptive}, listed in Table.~\ref{table 2}. 

\begin{table}[htbp]
    \caption{Action Set U}
    \centering
    \begin{tabular}{c | c c}
    \hline
    Action $u$ & $A(m/s^{2})$ & $\omega(rad/s)$\\
    \hline 
    Maintain $u_{1}$ & 0 & 0 \\
    Accelerate $u_{2}$ & 2 & 0 \\
    Decelerate $u_{3}$ & -2 & 0 \\
    Brake $u_{4}$ & -4 & 0 \\
    Turn left $u_{5}$ & 0 & $\pi/4$ \\
    Turn right $u_{6}$ & 0 & $-\pi/4$ \\
    \hline 
    \end{tabular}
    \label{table 2}
\end{table}

\noindent where $U$ is the action set, $u_{t}^{i}=[a_{t}^{i}, \omega_{t}^{i}]^T$ is the action of vehicle $i$ at $t$ frame, $a_{t}^{i}$ is acceleration, and $\omega_{t}^{i}$ is heading angle rate. Therefore, six feasible trajectories can be generated in each frame by constituting action $u_{t}^{i}$ with vehicle dynamics in Eq.~\ref{vehicle dynamics}.

\begin{equation}
f(s_{t}^{i},u_{t}^{i})=
\left[
\begin{aligned}
x_{t}^{i} & +v_{t}^{i}\cos(\gamma_{t}^{i})\Delta t \\
y_{t}^{i} & +v_{t}^{i}\sin(\gamma_{t}^{i})\Delta t \\
v_{t}^{i} & +a_{t}^{i}\Delta t \\
\gamma_{t}^{i} & +\omega_{t}^{i}\Delta t 
\label{vehicle dynamics}
\end{aligned}
\right]
\end{equation}
where $f$ denotes the uni-cycle model, $s_{t}^{i}=[x_{t}^{i},y_{t}^{i},v_{t}^{i},\gamma_{t}^{i}]^T$ denotes the state of a vehicle, $(x_{t}^{i},y_{t}^{i})$, $v_{t}^{i}$ and $\gamma_{t}^{i}$ are the position, speed, and yaw angle of vehicle $i$ at $t$ frame, respectively. 

Furthermore, in a non-cooperative game, a discreet trade-off between the consequence of each action and the rival's possible corresponding response will be made in every $\Delta t=0.1s$. Finally, after taking every action in Table~\ref{table 2} into account, the best response to cope with the rival’s each action can be concluded. After each player strives for the best response which results in the lowest cost or the highest reward, Nash equilibrium will be accomplished. In Nash equilibrium, none of the players are able to acquire a lower cost by unilaterally adjusting their actions. Therefore, the mathematical definition of Nash equilibrium is defined by

\begin{equation}
    R(u_{t}^{i^{*}},u_{t}^{-i^{*}}) \geq R(u_{t}^{i},u_{t}^{-i^{*}}) 
    \label{reward}
\end{equation}
where $R$ is the reward function, $u_{t}^{i^{*}}$ and $u_{t}^{-i^{*}}$ are the action performed by player $i$ and players other than player $i$ when game achieves Nash equilibrium at $t$ frame, $u_{t}^{i}$ is arbitrary action of player $i$ in action set $U$.

In some cases, Nash equilibrium is not always the optimal solution \cite{chremos2020game}. However, we do not regard this as a drawback, but as a proxy for the limitation of human reasoning and the consequence of being absolutely rational. As in the real-world, not all games have a happy ending.

Here we considered a reward consisting of efficiency, comfort, and safety, denoted by distance to the destination, offset to the expected path, and Time to Collision (TTC), respectively. Then, the reward of each possible action can be calculated by $s_{t}^{i}$, $u_{t}^{i}$, and $u_{t}^{-i}$  in Eq.~\ref{reward}.

\begin{equation}
    R(s_{t}^{i},u_{t}^{i},u_{t}^{-i})=[-d_{t}^{i}-o_{t}^{i}+\frac{1}{TTC_{t}^{i}}]\boldsymbol{\theta}^{T}
    \label{reward}
\end{equation}
where $d_{t}^{i}$ is the distance between the contemporary position of vehicle $i$ and its destination at $t$ frame, $o_{t}^{i}$ is the offset to the expected path, $TTC_{t}^{i}$ is the time to collision, and $\boldsymbol{\theta}^{T}$ is driver's decision-making preference weight.

Therefore, combining Eq.~\ref{vehicle dynamics} with Table ~\ref{table 2}, six feasible trajectories can be generated in each frame. Max entropy IRL was then introduced to quantify the difference between feasible trajectories and expert demonstrations to calibrate HVs different decision-making preferences.

\subsection {Decision-making Preference Calibration by IRL}
After dividing drivers into groups and generating feasible trajectories through non-cooperative game. IRL was introduced to excavate inherent characteristics that influence the expert decision. IRL was proposed later than behavior cloning. Though they share many similarities. Differing from simply imitating expert maneuvers, IRL tries to infer the reason why experts make their decision and then optimize the strategy. In other words, except directly learning the state-action mapping, IRL infers the form of reward weight and optimizes maneuvers through it.

Because of the aforementioned advantages, IRL has been widespread in many fields. Among many IRL algorithms, the maximum entropy method stands out due to its ability to address the ambiguity. A human driver follows the stochastic policies that may lead to a different distribution of candidate decisions \cite{huang2021driving}. The principle of maximum entropy was introduced to choose the distribution that does not exhibit any additional preference. Therefore, ambiguity was resolved. Generally, the probability of a trajectory is proportional to the exponential of the reward in Eq.~\ref{eq P}.

\begin{equation}
    P(\zeta|\boldsymbol{\theta})=\frac{1}{Z(\boldsymbol{\theta})}e^{R(\zeta)}=\frac{1}{Z(\boldsymbol{\theta})}e^{\boldsymbol{\theta}^{T}f_{\zeta}}
    \label{eq P}
\end{equation}
where $P(\zeta|\boldsymbol{\theta})$ is the probability of trajectory $\zeta$, $R(\zeta)$ is the corresponding reward function of expert trajectory $\zeta$ which is the multiplication of reward weight vector $\boldsymbol{\theta}^{T}$ and trajectory feature vector $f_{\zeta}$, and $Z(\boldsymbol{\theta})$ is the partition function.

When confronted with a continuous or high-dimensional space problem, the partition function may fail to converge. However, in a finite horizon problem, the reward weights in maximizing entropy are certain to be convergent \cite{ziebart2008maximum}. Therefore, limited feasible trajectories were generated and Eq.~\ref{eq P} could be rewritten as follows.

\begin{equation}
    P(\zeta|\boldsymbol{\theta})=\frac{e^{\boldsymbol{\theta}^{T}f_{\zeta}}}{\sum_{i=1}^{M}e^{\boldsymbol{\theta}^{T}f_{\Tilde{\zeta}}}}
    \label{eq P2}
\end{equation}
where $f_{\Tilde{\zeta}}$ is the feature of feasible trajectory $\Tilde{\zeta}$, and $M$ is the finite feasible trajectory. By this approximation, probability is much easier to calculate.

The kernel of IRL is to adjust the weights of reward function to yield a maneuver that matches with expert demonstrations. To fulfill that purpose, the likelihood of expert demonstrations should be maximized by adjusting $\boldsymbol{\theta}^{T}$.

\begin{equation}
    \max\limits_{\boldsymbol{\theta}}L(\boldsymbol{\theta}) = \max\limits_{\boldsymbol{\theta}}\sum\limits_{\zeta\in{D}}logP(\zeta|\boldsymbol{\theta})
    \label{eq max}
\end{equation}
where $L(\boldsymbol{\theta})$ is the likelihood function and also the objective function, $D=\{\zeta_{i}\}_{i=1}^{N}$ is the trajectory set containing $N$ expert demonstrations which are the real-world human trajectories we screened and clustered before.

Substituting $P(\zeta|\boldsymbol{\theta})$ in Eq.~\ref{eq max} with Eq.~\ref{eq P2}, $L(\boldsymbol{\theta})$ can be rewritten as

\begin{equation}
    L(\boldsymbol{\theta}) = \sum\limits_{\zeta\in{D}}
    [\boldsymbol{\theta}^{T}f_{\zeta}+log\sum_{i=1}^{M}e^{\boldsymbol{\theta}^{T}f_{\Tilde{\zeta}}}]
    \label{eq L}
\end{equation}

This function is convex and gradient-based optimization is used for solving optimal reward weight. The gradient can be written as the difference between the expert demonstrations $f_{\zeta}$ and the feasible trajectories $f_{\Tilde{\zeta}}$ in Eq.~\ref{eq grad L}.

\begin{equation}
    \nabla_{\boldsymbol{\theta}} L(\boldsymbol{\theta})=\sum\limits_{\zeta\in{D}}[f(\zeta)-\sum_{i=1}^{M}\frac{e^{\boldsymbol{\theta}^{T}f_{\zeta}}}{\sum_{i=1}^{M}e^{\boldsymbol{\theta}^{T}f_{\Tilde{\zeta}}}}]
    \label{eq grad L}
\end{equation}

L2 regularization is introduced to the likelihood function in order to prevent overfitting and $\lambda>0$ is the regularization parameter. Thus, Eq.~\ref{eq L} and Eq.~\ref{eq grad L} can be rewritten as

\begin{equation}
    L(\boldsymbol{\theta}) = \sum\limits_{\zeta\in{D}}
    [\boldsymbol{\theta}^{T}f_{\zeta}+log\sum_{i=1}^{M}e^{\boldsymbol{\theta}^{T}f_{\Tilde{\zeta}}}]-\lambda\boldsymbol{\theta}^{2}
    \label{eq L2}
\end{equation}

\begin{equation}
    \nabla_{\boldsymbol{\theta}} L(\boldsymbol{\theta})=\sum\limits_{\zeta\in{D}}[f(\zeta)-\sum_{i=1}^{M}\frac{e^{\boldsymbol{\theta}^{T}}f_{\zeta}}{\sum_{i=1}^{M}e^{\boldsymbol{\theta}^{T}}f_{\Tilde{\zeta}}}]-2\lambda\boldsymbol{\theta}
    \label{eq grad L2}
\end{equation}

Then, the pseudocode of maximum entropy IRL is summarized in Algorithm.~\ref{pseudocode}. Based on the clustering results, three groups of expert demonstrations were respectively trained by IRL to calibrate the preference of different drivers' decision. Feasible trajectories and corresponding features can be obtained by utilizing non-cooperative game and Eq.~\ref{reward}, respectively. 

% =======
% pseudocode 1
% =======
\begin{algorithm}
\caption{Maximum entropy inverse reinforcement learning.}\label{pseudocode}
\KwData{Three expert demonstrations dataset, learning rate $ \alpha$, regularization parameter $\lambda$, training epoch $E$}
\KwResult{Optimized reward weights $\boldsymbol{\theta}^{T}$, including efficiency, comfort, and safety}
Initialize $\boldsymbol{\theta}^{T} \leftarrow \mathcal{N}(0,0.05)$\;
Initialize buffer $B \leftarrow []$\;
Compute human features $f\leftarrow \sum_{i=1}^{M}f_{\zeta}$\;
\While{$\zeta_{i}\in D$}{Generate feasible trajectories $f_{\Tilde{\zeta}}$\;
Calculate features of $\Tilde{\zeta}$\;
Add features to buffer $B\stackrel{+}\leftarrow \Tilde{\zeta}$\;
$i \leftarrow i+1$\;}
\While{$epoch < E$}{Calculate the likelihood function of each sample in buffer $B$\;
$L(\boldsymbol{\theta}) = \sum\limits_{\zeta\in{D}}[\boldsymbol{\theta}^{T}f_{\zeta}+log\sum_{i=1}^{M}e^{\boldsymbol{\theta}^{T}f_{\Tilde{\zeta}}}]$\;
Calculate the gradient $\nabla_{\boldsymbol{\theta}} L(\boldsymbol{\theta})$\;
Update reward weights $\boldsymbol{\theta} \leftarrow \boldsymbol{\theta} + \alpha \nabla_{\boldsymbol{\theta}} L(\boldsymbol{\theta})$\;
$\boldsymbol{\theta}^{*} \leftarrow \boldsymbol{\theta}$\;}
\end{algorithm}

After training, the reward weights of different driver groups can be calibrated through maximizing entropy and iterating, shown in Table~\ref{table 3}.

% =======
% TABLE. 03
% =======
\begin{table}[htbp]
    \caption{IRL Calibration Results}
    \centering
    \begin{tabular}{c|c c c}
    \hline
     \multirow{2}{*}{Type}&  \multicolumn{3}{c}{Reward weights} \\
     \cline{2-4}
         & efficiency & comfort & safety\\
     \hline
     Aggressive HV & 8.33 & 1.56 & 3.69 \\
     Normal HV & 8.2 & 1.72 & 5.7 \\
     Conservative HV & 7.79 & 2.1 & 8.44 \\
    \hline
\end{tabular}
    \label{table 3}
\end{table}

As in Table~\ref{table 3}, the aggressive driver group possessed the highest efficiency value, and lowest safety and comfort value, indicating their preference for passing at a high speed and more willingness to detour rather than stop and wait. Meanwhile, conservative drivers were more concerned about driving comfort and safety, leading to conservative decisions. The reward weights of normal drivers were in the middle range, these drivers do not have an over preference and try to balance efficiency, comfort, and safety while driving. 

In conclusion, the above results were in line with real-world expectations. On the other hand, they also confirmed that our definition of the cluster centers in subsection A was reasonable. After calibrating different drivers’ preferences on decision making, the action corresponding to Nash equilibrium can be inferred and the next state of vehicle can be calculated by substituting this action into Eq.~\ref{vehicle dynamics}. 

So far, we have modeled the heterogeneous HVs for establishing a mixed environment. Next, we designed a cooperative driving framework that enables CAVs to not only interact, but also actively coordinate with other vehicles, advancing traffic efficiency and safety.

% === IV. Decision making and trajectory planning for CAVs=======================================
% =================================================================================
\section{Cooperative Decision Making and trajectory planning for CAVs}
Unlike the unpredictable and uncontrollable of HVs' trajectories, CAVs are capable of cooperating and even coordinating with others to nudge the system to greater efficiency. As shown in Fig.~\ref{framework}, the cooperative driving algorithm for CAVs is introduced in three subsections. Level-k game and cooperative game are introduced in subsection A and subsection B to search for an optimal k-allocation solution refers to the highest system efficiency. In subsection C, Lattice planner is used for planning an optimal and collision-free trajectory based on k-allocation results.

\subsection {Resolving Multiple Traffic Streams Conflict by Level-k Game}

Level-k game assumes that players generate strategies based on their depths of reasoning. In addition, k can also be defined as ROW \cite{wang2020performance}, driving style \cite{tian2018adaptive, tian2020game}. Here, considering the absence of traffic lights will cause multiple traffic streams to intersect in time and space, we denoted k as ROW to resolve conflicts.

In the case of trajectory planning, level-0 drivers regard other vehicles as static obstacles and plan their trajectory on this basic assumption. Then level-1 drivers will infer the planning trajectory of level-0 drivers and then plan their trajectories while avoiding collision with level-0 trajectory. By that analogy, all drivers may obtain the expected paths based on their depth of reasoning. Therefore, the expected path of vehicle $i$, denoted as $\zeta_{i}$, can be calculated from

\begin{equation}
    \zeta_{i}(k) = \max R(k, s^{i}_{u}, \zeta_{j}(k-1))
    \label{level k}
\end{equation}
where $\zeta_{i}(k)$ is the trajectory of vehicle $i$ with $k$ level, and the trajectory is determined by maximizing the reward function $R$.

In this paper, our CAV is free to generate its depths of reasoning to maximize the system's overall efficiency. In addition, when multiple traffic streams intersect, a unique k will be assigned to each stream to prevent collision or deadlock. Researchers also pointed out that normally reasoning depths of human are less than or equal to 2 according to \cite{tian2020game, wang2022comprehensive}. Therefore, to prevent CAVs from behaving unreasonably, traffic streams with k greater than 2 were corrected to k equal to 2 based on the limited human reasoning depths.

Level-k game provided a theoretical basis for the orderly passage of vehicles. Though unsignalized intersections failed to resolve conflict from the temporal dimension, countable streams lead to finite k-allocation solutions. By optimizing the k-allocation of streams, system optimum can be achieved and the conflict can be resolved. Therefore, cooperative game was utilized to quantify the system performance of each k-allocation solution.

\subsection {Searching Best k-allocation through Cooperative Game}
Reservation-based control methods such as First Come First Serve (FCFS) and Batch-strategy have a long history and are widely used due to their succinct form \cite{yu2019managing}. FCFS control simply assumes that ROW is proportionate to the entry order and firstly served as a reservation mechanism for autonomous vehicles ascend to \citet{dresner2008multiagent}. Batch-strategy believed that intersections will be more efficient from a capacity of view if served in batches. Therefore, vehicles in each stream are grouped into batches and served at FCFS principle then. While these methods reckon without system efficiency and organize ROW merely depend on their order of entry.

Therefore,  cooperative game was combined with the aforementioned level-k game to optimize system efficiency. Differing from non-cooperative game that regards other participants as rivals and emphasizes maximizing individual profit. Cooperative game refers to participants uniting as several coalitions and trying to maximize system goals through collaborating with other coalitions. Meanwhile, cooperative game should meet the requirement of superadditivity, individual rationality, and group rationality \cite{yang2018cooperative, xing2022secure}. 

Superadditivity refers to the inequality of the reward of coalitions and the total reward of separate individual vehicles. It can be expressed by

\begin{equation}
    R(V_{i})+R(V_{j})\leq R(V_{i}\cup V_{j})
    \label{superadditivity}
\end{equation}
where $R(V_{i})$ is the reward function of vehicle $i$, and $V_{i}\cup V_{j}$ stands for the coalitions consists of vehicle $i$ and vehicle $j$. In addition, individual rationality should satisfy the following inequality.

\begin{equation}
    R(V_{i})\leq R_{i}(V_{i}\cup\ V_{j}...\cup V_{n})
    \label{individual rationality}
\end{equation}
where $R_{i}(V_{i}\cup V_{j}...\cup V_{n})$ represents the individual reward of vehicle $i$ under cooperative game that $n$ vehicles are united as a coalition. This inequality stipulates that each vehicle could obtain a higher reward when controlled with cooperative game than before. Finally, group rationality means that the system goals are the summation of individual rewards, i.e.,

\begin{equation}
    O=\sum\limits_{i\in N}R(V_{i})
    \label{group rationality}
\end{equation}
where $O$ is the system goal of a cooperative game and the sum of individual reward at the same time. System goals are often in the form of minimum overall delay or maximum efficiency such as moving faster or further. 

Through experiments, we finally chose the inverse distance to the destination to represent system efficiency as it has shown a better performance than other surrogate variables. Therefore, the objective function $O$ is as follows.

\begin{equation}
    O(k)=\max\sum\limits_{i\in lane}\sum\limits_{j\in group}\frac{1}{\overline{d}_{i,j}}
    \label{obj function}
\end{equation}
where O(k) is the value of objective function under a certain k-allocation solution, $\overline{d}_{i,j}$ is the average distance to the destination of the expected path belonging to vehicles in $i$ lane and $j$ group. Each k-allocation solution will lead to a different $\overline{d}_{i,j}$. 

It should be noted that this distance is not the current distance to destination but the average distance to destination of the planning trajectory under k-level. The value of $\overline{d}_{i,j}$ can be achieved by giving a specific k to the Lattice planner, which will be discussed in the next subsection.

At the same time, we were aware that going straight vehicles and turning vehicles have different impacts on the system. Therefore, coefficients of distance to the destination should be various and Eq.~\ref{obj function} can be rewritten as

\begin{equation}
\begin{split}
    \max(\sum\limits_{j\in group}\frac{1}{\overline{d}_{gs,j}} +\sum\limits_{j^{'}\in group^{'}}\frac{\alpha}{\overline{d}_{turn,j^{'}}})
    \label{obj function2}
\end{split}
\end{equation}
where $\alpha$ represents the weight of turning vehicles relative to going straight vehicles, $d_{gs}$ and $d_{turn}$ are the distance to destination of going straight vehicle and turning vehicle, respectively. Then, an investigation was carried out to unify the dimension of $\overline{d}_{gs}$ and $\overline{d}_{turn}$.

\begin{figure}
  \begin{center}
  \centerline{\includegraphics[width=2.5in]{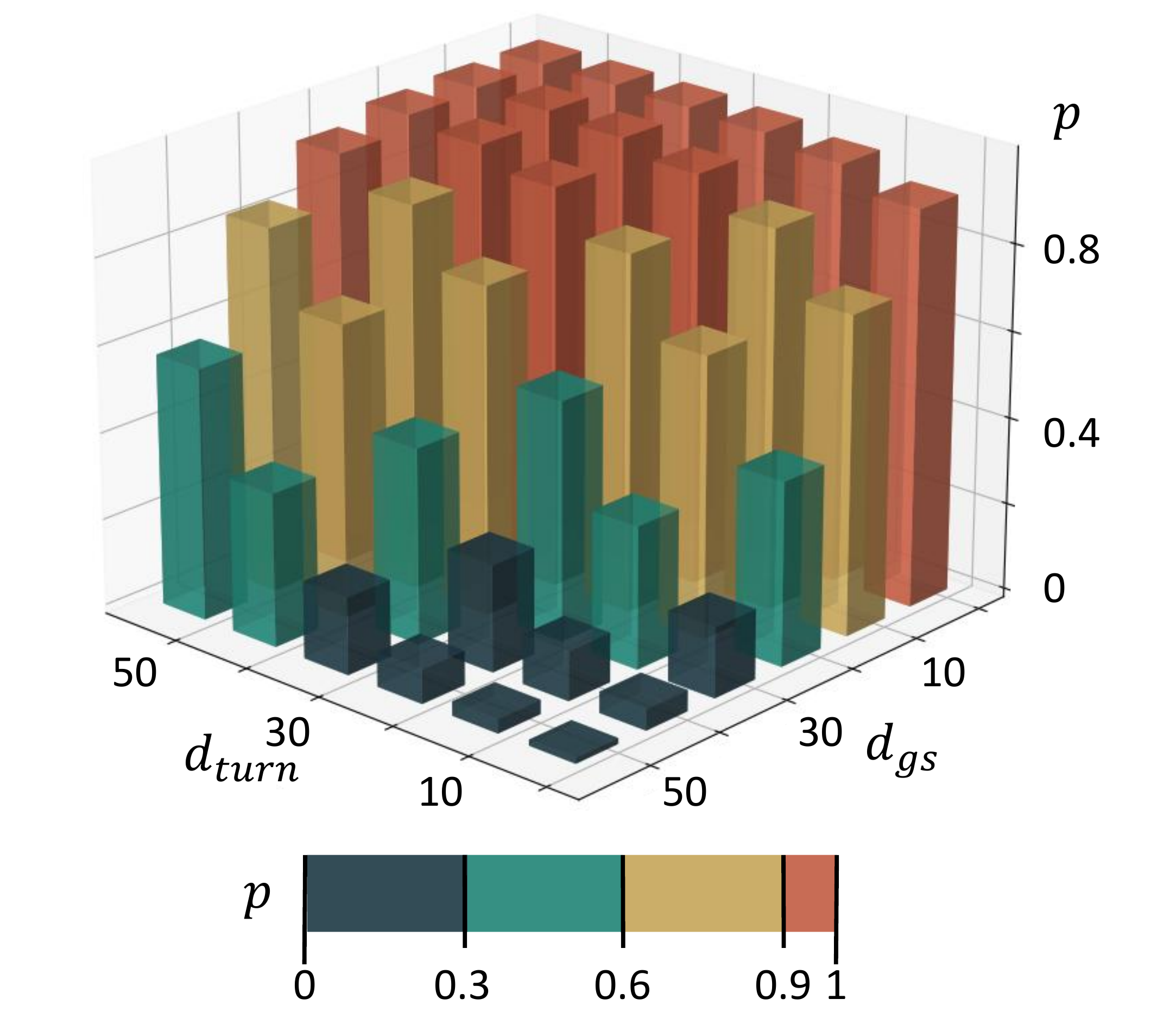}}
  \caption{Logistic relation of distance to destination and probability of going straight vehicle pass first.}\label{logit figure}
  \end{center}
\end{figure}

For the purpose of investigating the real-world relationship between going straight vehicles and turning vehicles, the interactions between going straight vehicles and turning vehicles were further extracted based on the XXJH data. Next, a logistic model was introduced to explore the relation between initial distance to the destination and the actual order of passage. After calibrating by interaction data, results are presented as follows.

\begin{equation}
    ln\frac{p}{1-p}=0.0925d_{turn}-0.1332d_{gs}+2.35
    \label{logit}
\end{equation}
where $p$ is the probability of going straight vehicle pass first. The relationship between $p$ and $d_{turn}$, $d_{gs}$ can be shown in Fig.~\ref{logit figure}. 

% =======
% FIG. 04
% =======

Substituting $p=0.5$ into Eq.~\ref{logit}, it yields that

\begin{equation}
    d_{turn}=1.44d_{gs}-25.4
    \label{logit3}
\end{equation}

Through the coefficients of $d_{turn}$ and $d_{gs}$, a turning vehicle will cost more to maintain equality with a going straight vehicle. This conclusion is consistent with relevant law which stipulated that turning vehicles should give precedence to going straight vehicles when in conflict at an intersection. In addition, by substituting Eq.~\ref{obj function2} into Eq.~\ref{logit3}, the objective function is derived as follows.

\begin{equation}
\begin{split}
    \max(\sum\limits_{j\in group}\frac{1}{\overline{d}_{gs,j}} +\sum\limits_{j^{'}\in group^{'}}\frac{1}{1.44\overline{d}_{gs,j^{'}}-25.4})
    \label{obj function3}
\end{split}
\end{equation}

Through the logistic model, the impacts of going straight vehicles and turning vehicles to the system were normalized. Social norms that going straight vehicle possess higher priority than turning vehicle at the intersection were also taken into consideration subtly during this normalization. Here, we named it normalized cooperative game.

As indicated above, in cooperative game, vehicles will unite as several coalitions. While maximizing system's overall efficiency is the same goal that all coalitions share. This process can be achieved by adjusting one’s strategy to influence other coalitions. Differing from the above non-cooperative game where influence has been seen as a discrete action-action mapping. A trajectory-trajectory strategy mapping was needed for cooperative game because of the form of the objective function. To calculate Eq.~\ref{obj function3}, each vehicle's planning trajectory should be inferred according to the concept of level-k game and Lattice planner.

Therefore, by combining normalized cooperative game with level-k game, k-allocation solution that led to optimal system efficiency can be easily obtained by enumerating finite k-allocation solutions. In addition, a trajectory planner was needed to generate trajectories based on the concept of level-k game to evaluate each k-allocation solution's performance. Hence the Lattice planner was introduced.

\subsection {Planning Optimal Trajectory with Lattice Planner}
Lattice planner is a well-accepted motion planner to generate optimal and collision-free trajectory according to expect path and obstacles. It mainly consists of coordinates transforming, sampling, and curvature polynomials fitting.

In the coordinates transforming step, the corresponding state in Frenet coordinates was transformed according to the initial state in Cartesian coordinates. Then, planning horizons and maximum steering angles were sampled. For improving computation speed, initial sampling space was relatively small. When it comes to no available solution, sampling space will be enlarged. Simulation will be shut down and print out no solution if there are still no available solutions. When it comes to multiple solutions, optimal trajectory was determined by reward. Finally, after sampling, curvature polynomials were used to generate the entire trajectory between the start point and end point. 

To control variables, the composition of rewards and reward weights for CAV was designed to be consistent with normal HV. The difference is HVs made decisions by searching Nash equilibrium in non-cooperative game and then generated trajectory through vehicle dynamics while CAVs replaced these steps with the NCL game we proposed and Lattice planner respectively. Algorithm.~\ref{CAV algorithm} describes the procedure of CAVs' cooperative decision making and trajectory planning.

% =======
% pseudocode 2
% =======
\begin{algorithm}
\caption{Cooperative decision making and trajectory planning for CAVs.}\label{CAV algorithm}
\KwIn{Vehicle's state $s_{t}^{i}$, traffic stream number $lane$, number of vehicles in each stream $group$}
\KwOut{Vehicle decision-making depths $k$ and next state $s_{t+1}^{i}$ according to planning trajectory}
Calculate all possible k-allocation solution $A \leftarrow A_{lane}^{2}$\;
Initialize objective function list of cooperative game $O$\;
\ForEach{$a\in A$}{Evaluate each $a$ performance\;
    \ForEach{$stream \in lane$}{\ForEach{$i \in group$}{Calculate average distance to destination $\overline{d}_{j}$ through $Lattice(k)$\;
    Normalize $\overline{d}_{gs,j} \leftarrow \overline{d}_{j}$\;
    Calculate objective function of this solution $O_{s} += \overline{d}_{gs,j}$\;}
    } 
    Add to $O \stackrel{+}\leftarrow O_{s}$
}
Find the best k-allocation solution $\max O$\;
Get each vehicle $k$ and plan its next state $s_{t+1}^{i}$ through $Lattice(k)$ and best k-allocation solution\;
\end{algorithm}

Steps in Algorithm~\ref{CAV algorithm} show the whole process of CAV cooperative decision making and trajectory planning. It will repeat every $\Delta t=0.1s$ until the simulation duration is reached. After finishing the modeling of HVs and CAVs, several experiments were conducted to validate our cooperative driving framework.

% === V. Experiment validation ========================================
% =================================================================================
\section{Simulation Validation}

\subsection {Simulation Design}
\begin{figure}
  \begin{center}
  \centerline{\includegraphics[width=3.8in]{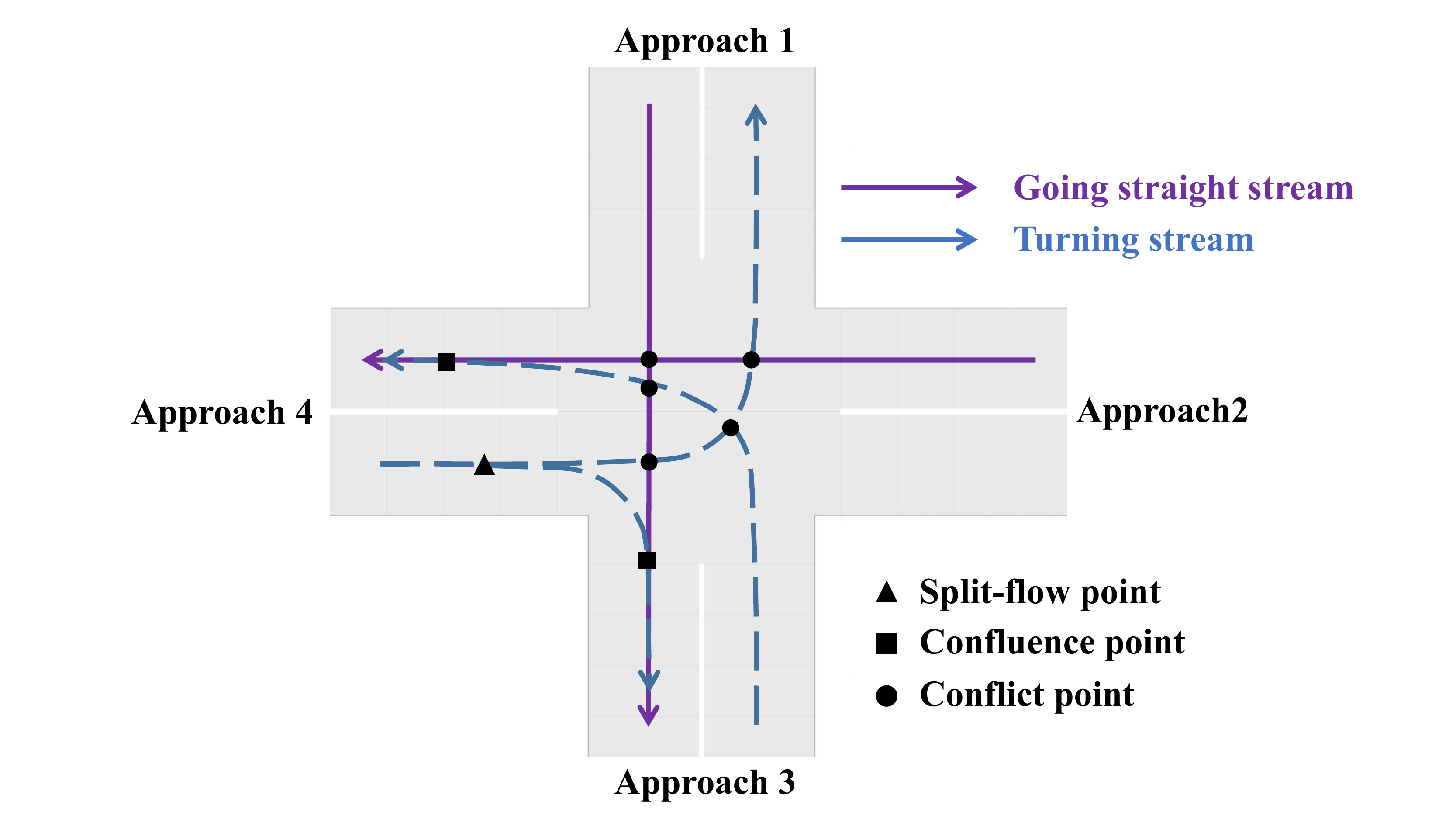}}
  \caption{Simulation environment.}\label{simulation}
  \end{center}
\end{figure}

To evaluate the performance of the proposed algorithm framework, an isolated four-approach unsignalized intersection in Fig.~\ref{simulation} was used as simulation environment. All approaches were 40 meters long and each approach contained one or two traffic streams as in Fig.~\ref{simulation}.

% =======
% FIG. 05
% =======

Although we simplified the number of traffic streams, nearly ten conflicts still exist according to the topological relationship. Except for various conflict points, simulation environment also included confluence point and split-flow point. This resulted in merging and diverging behaviors which further produced substantial impacts on vehicle behavior in mixed traffic \cite{guo2020merging}. Together with high-density traffic flow and heterogeneous HVs, we believe this unsignalized intersection is complicated enough for examining our algorithm framework.

In addition, preferences affect not only the drivers’ decisions,  but also their initial speed and target speed. Therefore, according to our prior knowledge based on XXJH data, HVs' initial speed was designed same as the average speed, and target speed same as the maximum speed. For example, according to Table~\ref{table 1}, an aggressive driver will enter the simulation environment at speed of 6.3m/s and expect to travel at speed of 7m/s if possible. In order to examine the effectiveness and robustness of our framework, three simulation cases were conducted in this paper. 

First of all, we compared NCL game and CL game (NCL game without normalization) with reservation-based control methods under different traffic densities to evaluate the rock-bottom control method we proposed. FCFS and Batch-strategy were used as the representative of reservation-based method based on \cite{yu2019managing}. Next, in order to verify the effectiveness of the proposed cooperative driving algorithm, simulation was conducted under different CAV rates of penetration. Finally, differing from regarding all human drivers make normal decisions, we introduced drivers with different personal preferences to our simulation environment based on IRL calibration results in Table~\ref{table 3}. 

All simulations lasted 2 minutes, which is 1200 frames. After simulation, trajectories of all vehicles were recorded for further evaluation.

\subsection {Case1: Comparison of Different Control Methods}
For the purpose of evaluating the results of different control methods. Simulations were conducted in a full CAV environment at first. Average travel speed was applied to evaluate system efficiency.

\begin{figure}
  \begin{center}
  \centerline{\includegraphics[width=3.8in]{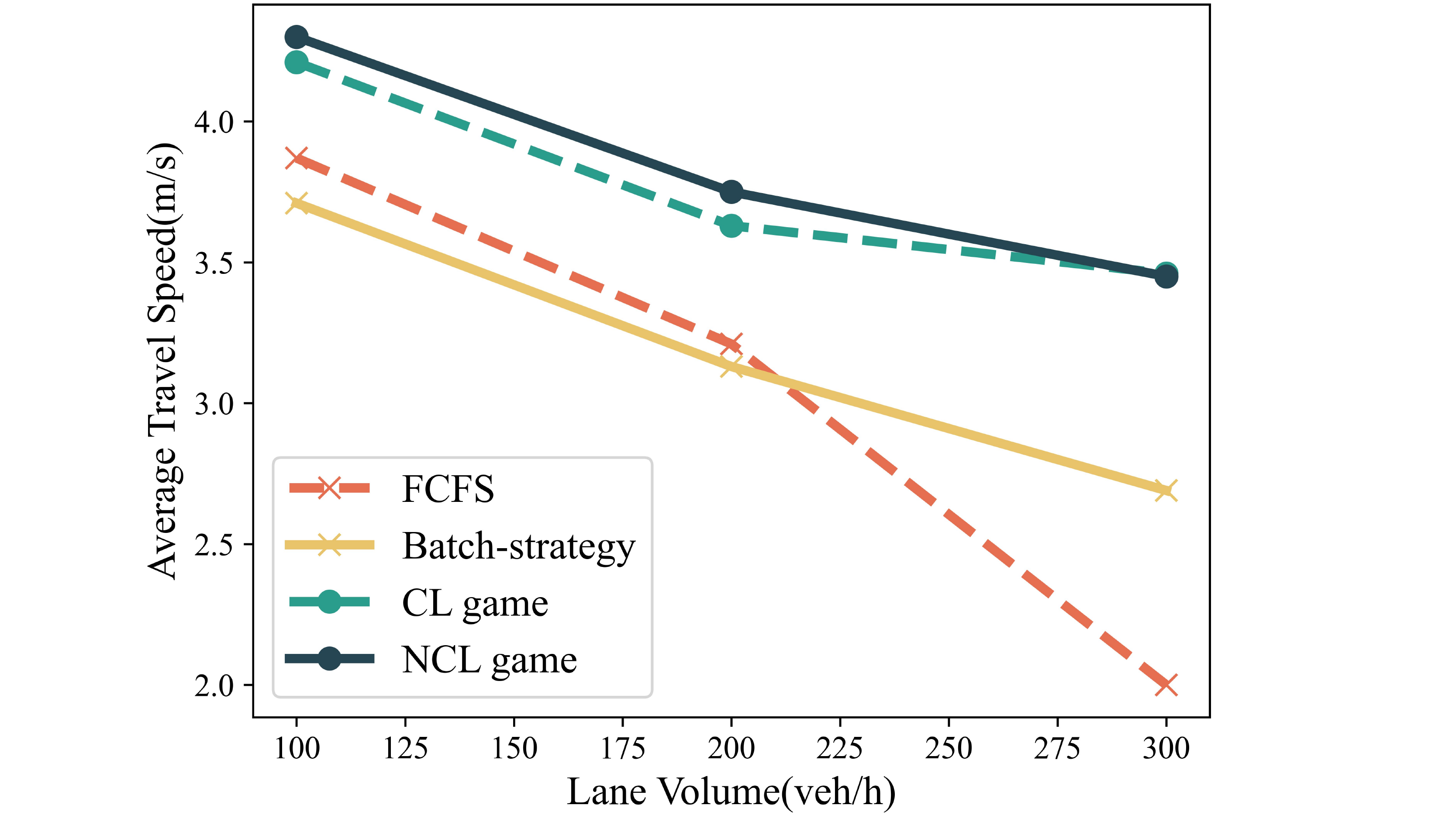}}
  \caption{Average travel speed comparison of control method under different lane volume.}\label{experiment1}
  \end{center}
\end{figure}

Fig.~\ref{experiment1} shows the average travel speed under different lane volumes. Clearly, it presented a negative relationship with lane volume, which is consistent with basic cognizance about the relationship of speed and volume according to \citet{greenshields1935study}. In addition, travel speed decreased dramatically under FCFS control when lane volume increased. But when lane volume was relatively low (under 200veh/h), FCFS control even outperform Batch-strategy because grouping vehicles into batches in low density may add unnecessary steps that lead to low efficiency.

% =======
% FIG. 06
% =======

As seen in Fig.~\ref{experiment1}, CL game and NCL game showed a higher travel speed than reservation-based control method at all lane volumes. Proving that the combination of cooperative game and level-k game is effective from the system point of view. Also, the gap between CL game and NCL game narrowed as lane volume rose. This phenomenon attributed to the number of vehicles in conflicting streams became more significant than the types of streams (turning stream or going straight stream). Therefore, with the increase in lane volume, the effect of normalization was subdued.

\begin{figure}
  \begin{center}
  \centerline{\includegraphics[width=3.8in]{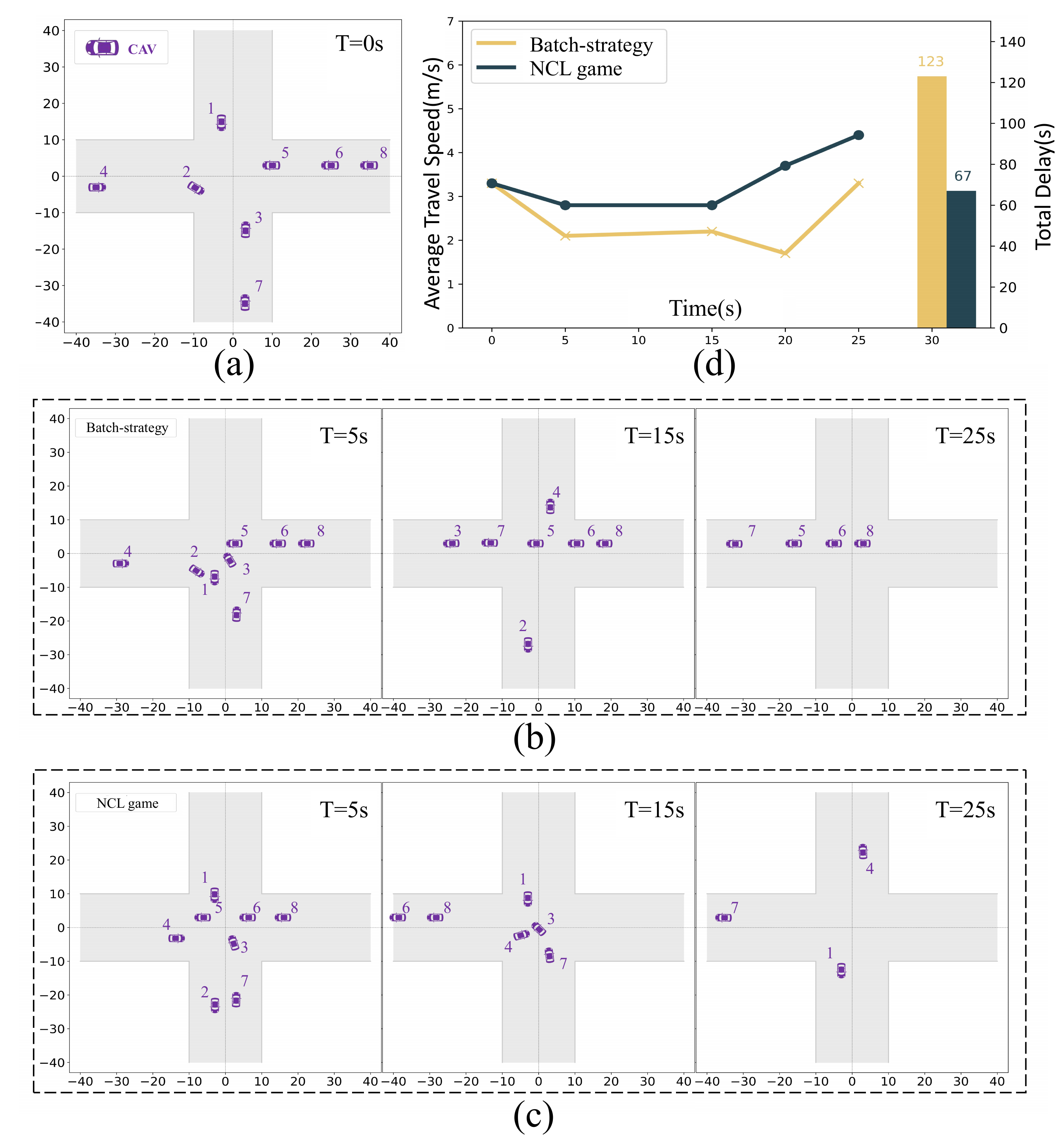}}
  \caption{Interaction case: comparison of NCL game and Batch-strategy: (a) shows the initial position of the vehicles; (b)-(c) show three subsequent steps of CAVs control with Batch-strategy and NCL game, respectively. (d) summarizes the average travel speed and total delay of this interaction case; numbers next to the vehicles refer to the order in which they enter the simulation environment.}\label{experiment1-8car}
  \end{center}
\end{figure}

Furthermore, Fig.~\ref{experiment1-8car} shows a case that eight vehicles exist in the simulation environment at the same time under Batch-strategy control and NCL game. Fig.~\ref{experiment1-8car} (a) depicts their initial position and the numbers beside vehicles stand for the enter sequence of a vehicle. Fig.~\ref{experiment1-8car} (b)-(c) exhibits the process of vehicles' position iterated over time. Owing to Batch-strategy serving vehicles as batches, $V_{7}$ (stands for vehicle 7 in the figure) obtained the same ROW as $V_{3}$ and passed in a queue. However, Batch-strategy still follows the principle of FCFS which led to $V_{1}$ passed first even if there was only one vehicle in its stream. This phenomenon has caused other streams which had conflict with $V_{1}$ have to recede no matter how many vehicles were in that group, resulting in inefficiency.

% =======
% FIG. 07
% =======

There is a clear trend that when controlled with NCL game, vehicles in Approach 2 possessed the highest ROW, and vehicles in Approach 3 were the second after screening all possible k-allocation solutions and finding the most efficient one. Fig.~\ref{experiment1-8car} (d) shows the average travel speed of the whole system at different simulation times. Obviously, NCL game we proposed brought a higher efficiency than Batch-strategy. In addition, the total delay of the whole intersection is almost half of controlled by Batch-strategy. It should be noted that this gap will ulteriorly amplify if vehicles ceaselessly enter the simulation environment.

Therefore, considering that high-density continuous traffic flow causes more trouble for CAV trajectory planning and more challenge to the whole system. In the following experiments, we fixed lane volume at 300 veh/h to evaluate the suitability of our work under mixed traffic.

\subsection {Case2: Experiment under Different ROP}
CAVs are expected to operate in traffic with HVs long into the future. Therefore, interactions between CAV and HV should be fully studied under different circumstances. Building on the experiment above, we evaluated the intersection system operation status under different ROP with NCL game and Batch-strategy as control methods. Each control method was conducted under three ROP, i.e., 100\%, 60\%, and 20\%.

\begin{figure}
  \begin{center}
  \centerline{\includegraphics[width=3.6in]{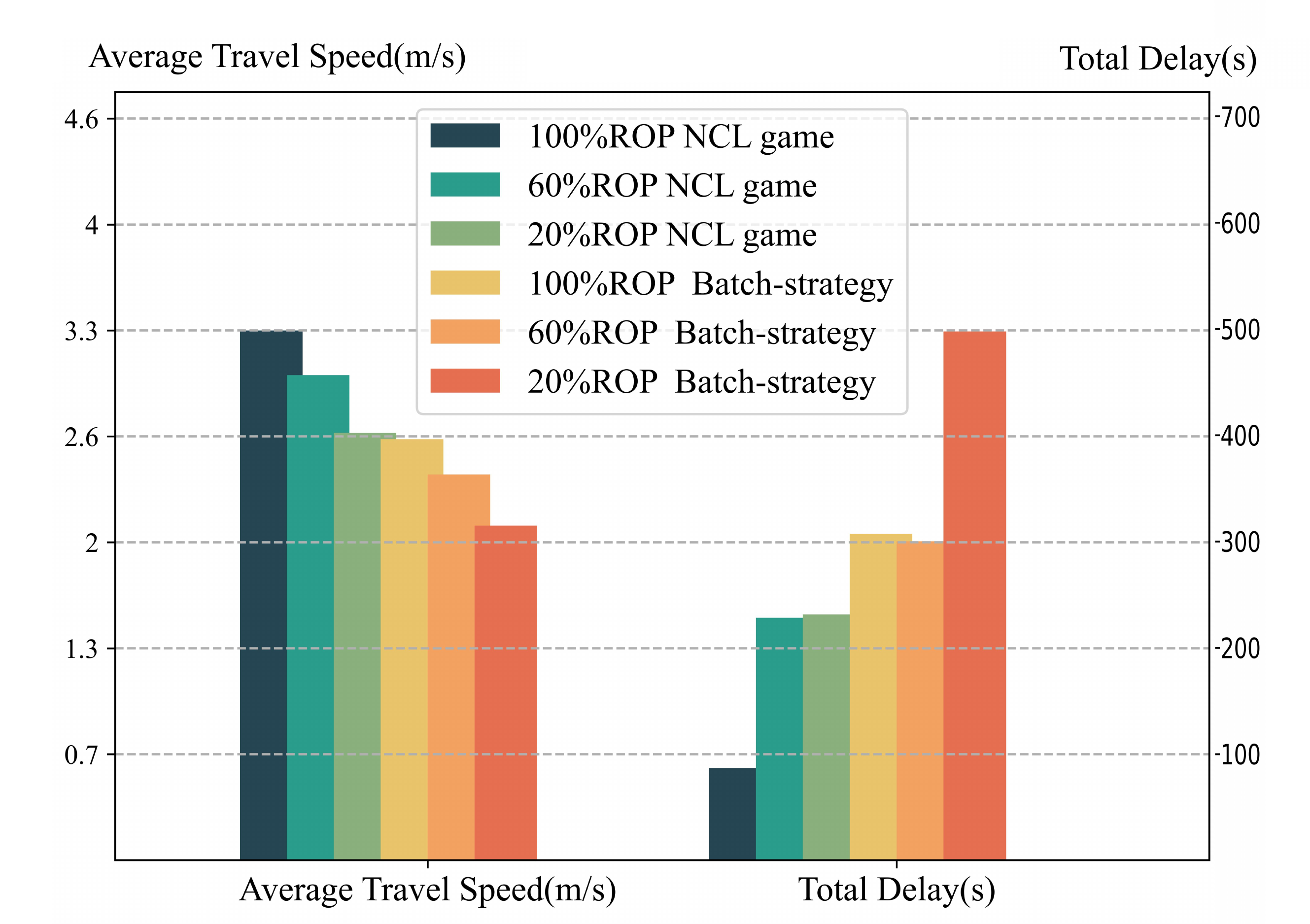}}
  \caption{Average travel speed and delay under different ROP.}\label{experiment2-v}
  \end{center}
\end{figure}

Average travel speed and total delay were calculated and standardized in Fig.~\ref{experiment2-v}. There is a clear trend when ROP rose, average travel speed increased and total delay decreased, representing the improvement of system efficiency. In addition, when CAV was controlled with NCL game, system showed better performance at all ROP. From another perspective, the minimum average travel speed when using NCL game control was even higher than the maximum travel speed using Batch-strategy control which confirmed that high-density continuous traffic flow indeed amplified the gap between NCL game and Batch-strategy. Same conclusions can be applied to total delay, proving that NCL game we proposed is superior to the traditional reservation-based method and is suitable for the mixed traffic flow containing both CAVs and HVs.

% =======
% FIG. 08
% =======

\begin{figure}
  \begin{center}
  \centerline{\includegraphics[width=3.4in]{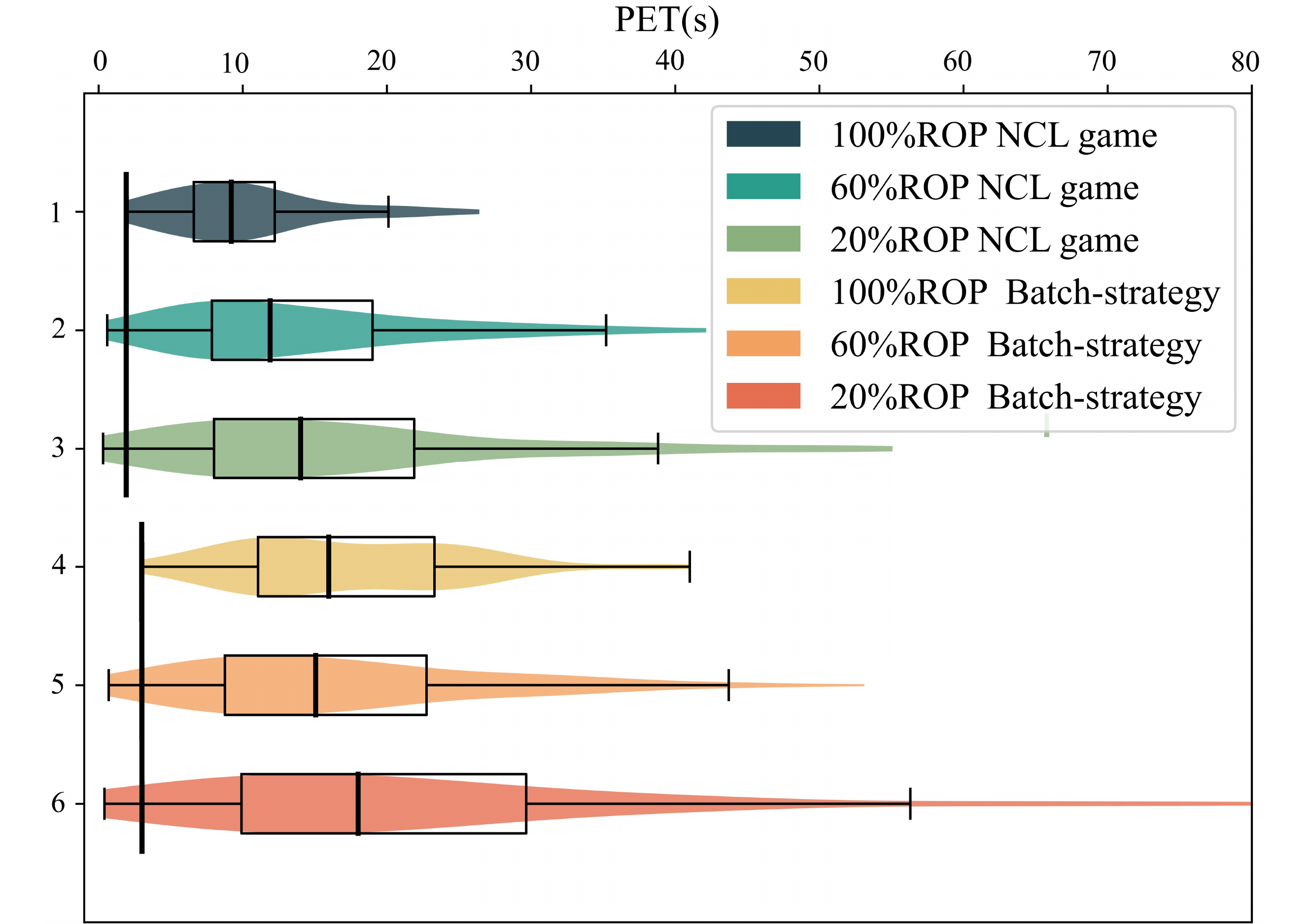}}
  \caption{Distribution of PET under different ROP.}\label{experiment2-pet}
  \end{center}
\end{figure}

The above analysis was aimed at evaluating the efficiency of intersection. Safety should also be noted because of its dependency on people’s trust in CAV which is currently a major obstacle to the popularization of CAV. Hence the PET was introduced to evaluate our algorithm from a safety perspective. Initially, PET was introduced by \citet{allen1978analysis} as the time between the moment when the front vehicle leaves conflict area or invasion line and the rear vehicle reaches conflict area or invasion line. The distribution of PET under different ROP and control methods is shown in Fig.~\ref{experiment2-pet}.

% =======
% FIG. 09
% =======

According to the box plots above, the distribution of PET became more discrete when ROP decreased, which indicates that interactions between vehicles were more chaotic and inefficient. At the same time, the decline of ROP also led to higher maximum PET and average PET, which symbolizes a longer interval between vehicle's passage based on the concept of PET. In other words, as the proportion of HV rose, interactions became conservative and hazardous. Thus, deadlocks were more likely to occur. 

Another line of evidence came from minimum PET, which indicates the most dangerous interaction case during the whole simulation duration. Fig.~\ref{experiment2-pet} shows a positive correlation between ROP and minimum PET, proving that increasing the number of our CAV is of help to improve system safety.

\begin{figure}
  \begin{center}
  \centerline{\includegraphics[width=4in]{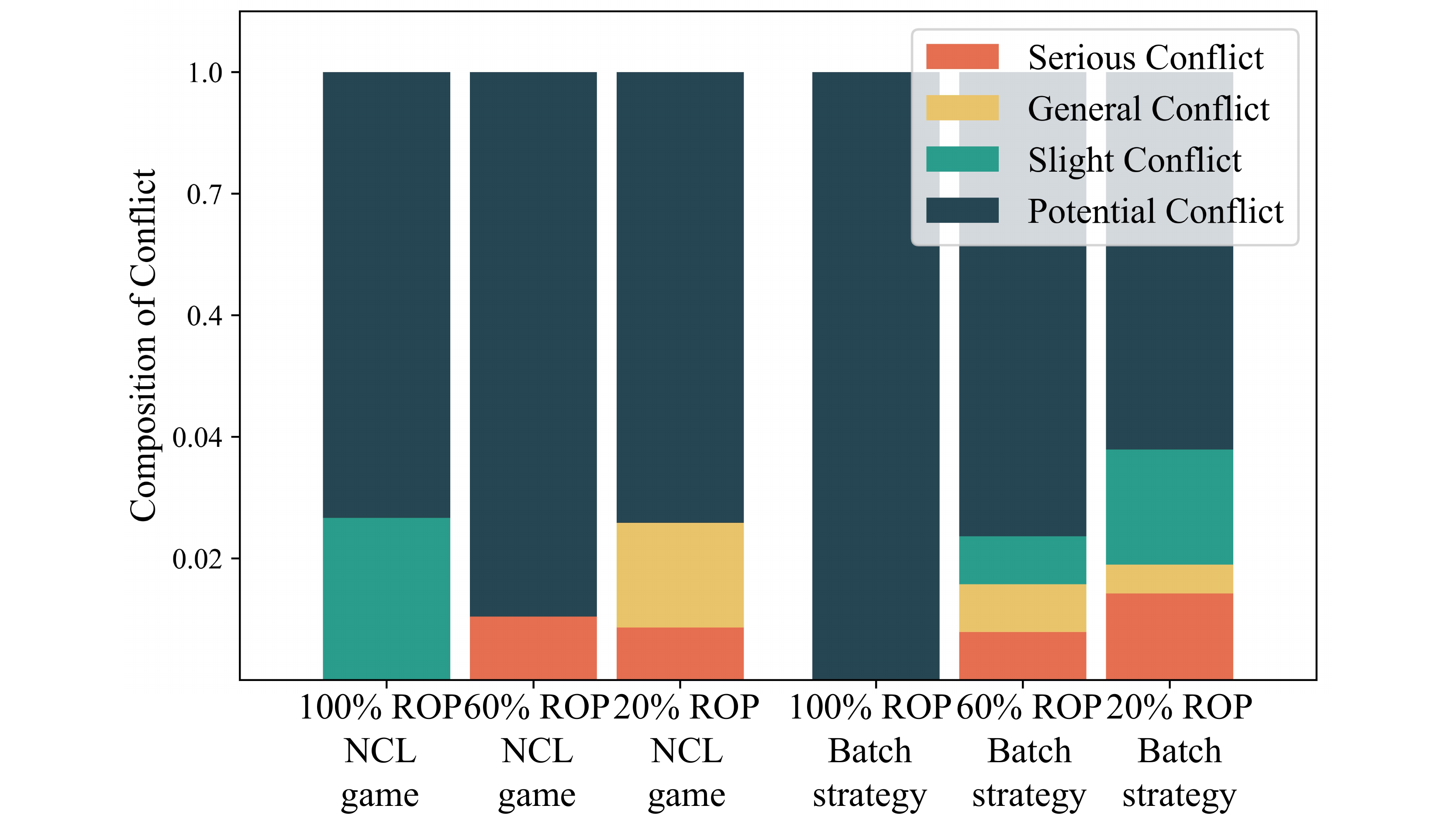}}
  \caption{Composition of conflict under different ROP.}\label{experiment2-compositionpet}
  \end{center}
\end{figure}

\citet{qi2020modified} determined an appropriate threshold of PET from conflict data. Dividing conflict into four sections. PET$\textless0.7s$ means a serious conflict, $0.7s\leq$PET$\textless1.31s$ means a general conflict, $1.31s\leq$PET$\textless2.25s$ means a slight conflict, and PET$\geq2.25s$ means a potential conflict. Furthermore, we calculated and visualized the conflict composition in Fig.~\ref{experiment2-compositionpet} based on above partition.

% =======
% FIG. 10
% =======

Fig.~\ref{experiment2-compositionpet} shows that interaction was either slight conflict or potential conflict in all CAV environment (100\% ROP). What is more, the proportion of serious conflict and general conflict  increased with the decrease of ROP. 

Through average travel speed, total delay, and PET, we have proven that the CAV we designed has the ability to nudge the intersection system toward greater efficiency and safety. However, in addition to high-density and mixed traffic, human drivers possess different decision-making preferences that leads to heterogeneous traffic. Therefore, interaction with heterogeneous HVs involved will fortify the persuasion of our CAV algorithm framework.

\subsection {Case3: Heterogeneous Human Driver Involved}
\begin{figure}
  \begin{center}
  \centerline{\includegraphics[width=3.8in]{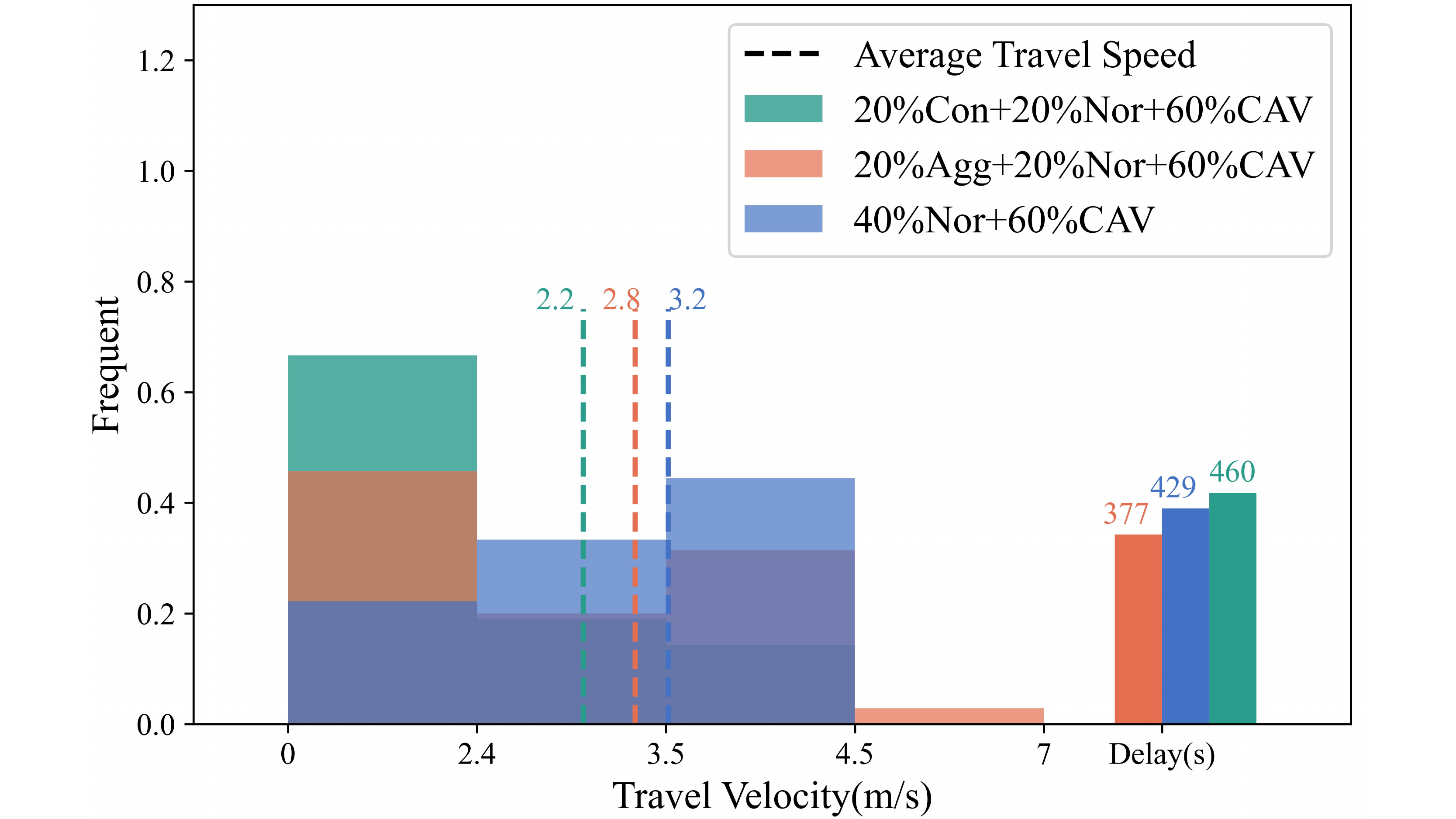}}
  \caption{Average travel speed and total delay under different vehicle compositions.}\label{experiment3-agg}
  \end{center}
\end{figure}

We conducted three experiments at 60\% ROP at first. Instead of generating all HVs make normal decisions while interacting with other vehicles like in Case 2. Half of HVs were replaced by aggressive type or conservative type, which may produce more unpredictable behavior. Total delay and the distribution of average travel speed are shown in Fig.~\ref{experiment3-agg}.

% =======
% FIG. 11
% =======

Fig.~\ref{experiment3-agg} shows that when there is only normal type HV mixed with CAVs in simulation environment, average travel speed was significantly higher than aggressive HV or conservative HV involved. In addition, when 20\% of normal HV was replaced by aggressive HV, though aggressive HV has a higher initial speed and target speed as we appointed in subsection $A$, the proportion of low-speed (velocity $\textless 2.4m/s$) vehicles increased on the contrary. Not to mention that the proportion of low-speed vehicles will increase sharply from 22\% to 67\% if 20\% normal vehicles were replaced by 20\% conservative vehicles. Though little difference was found in terms of total delay between the three experiments, the above experiments yet emphasized the difficulty when interacting with heterogeneous HVs whose decisions are fickle. 

Whereas these findings only prove our CAV is capable of interacting with heterogeneous HVs because no collision or deadlock happened, experiments should be carried out to examine if our CAV can improve system efficiency to testify their ability of coordinating with heterogeneous HVs. Therefore, a real-world based environment was established according to the HV composition in XXJH data. Finally, the experimental group consisted of 13\% aggressive HV, 41\% normal HV, and 46\% conservative HV. In the control group, normal HVs were replaced by the CAV we designed (note that they share the same initial speed, target speed, reward composition, and reward weights).

% =======
% FIG. 12
% =======
\begin{figure}
  \begin{center}
  \centerline{\includegraphics[width=3.8in]{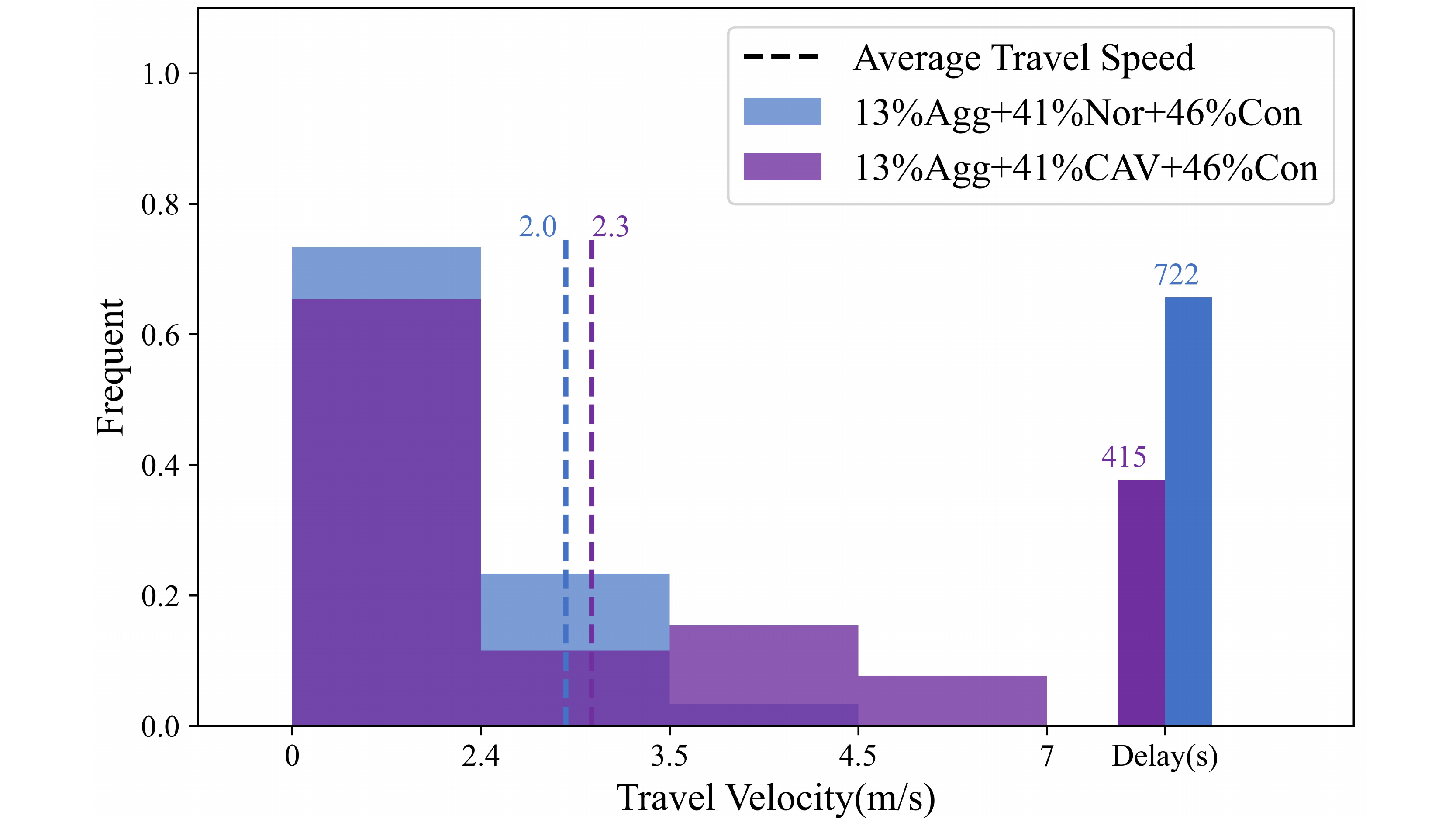}}
  \caption{Average travel speed and total delay based on XXJH vehicle composition.}\label{experiment3-real}
  \end{center}
\end{figure}

% =======
% FIG. 13
% =======
\begin{figure}
  \begin{center}
  \centerline{\includegraphics[width=3.4in]{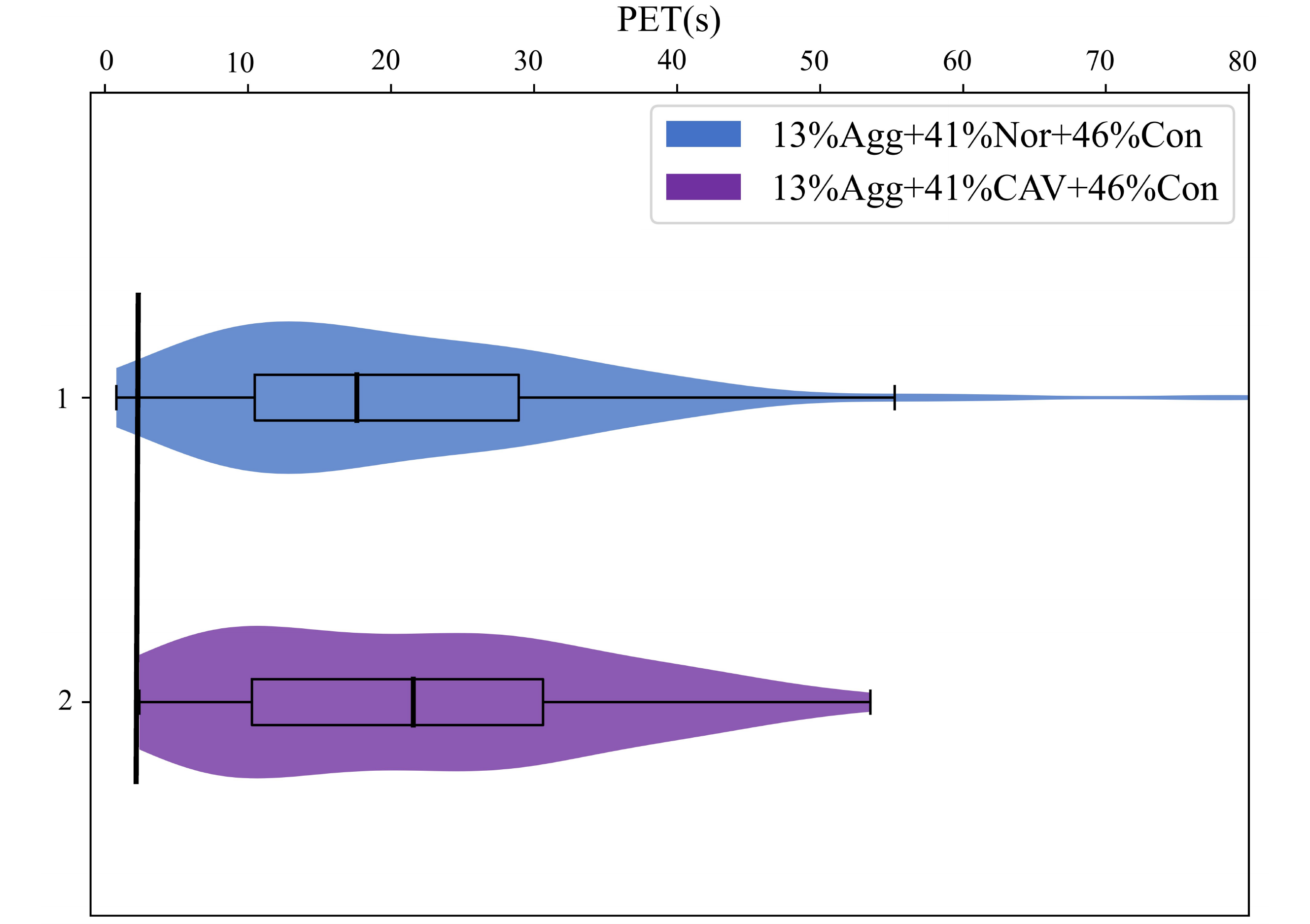}}
  \caption{Distribution of PET based on XXJH vehicle composition.}\label{experiment3-pet}
  \end{center}
\end{figure}

According to Fig.~\ref{experiment3-real}, after replacing normal HV with our CAV, system has been improved in all aspects. Specifically, average travel speed of the intersection system increased by 0.3m/s and the proportion of high-speed vehicles increased. The more striking change was a 42.5\% reduction in total delay which confirmed the robustness and the superiority of our CAV algorithm framework under a heterogeneous mixed environment.

Similar to Case2, we calculated the PET of the interactions during the simulation duration. As seen in Fig.~\ref{experiment3-pet}, after replacing normal HV with CAV, distribution of PET became more concentrated and minimum PET increased, indicating that our CAV has the ability to properly organize other vehicles' passage and improve driving safety. We further carried a interaction case to visualize the different system evolutionary processes of normal HV and CAV involved.

% =======
% FIG. 14
% =======
\begin{figure}
  \begin{center}
  \centerline{\includegraphics[width=3.5in]{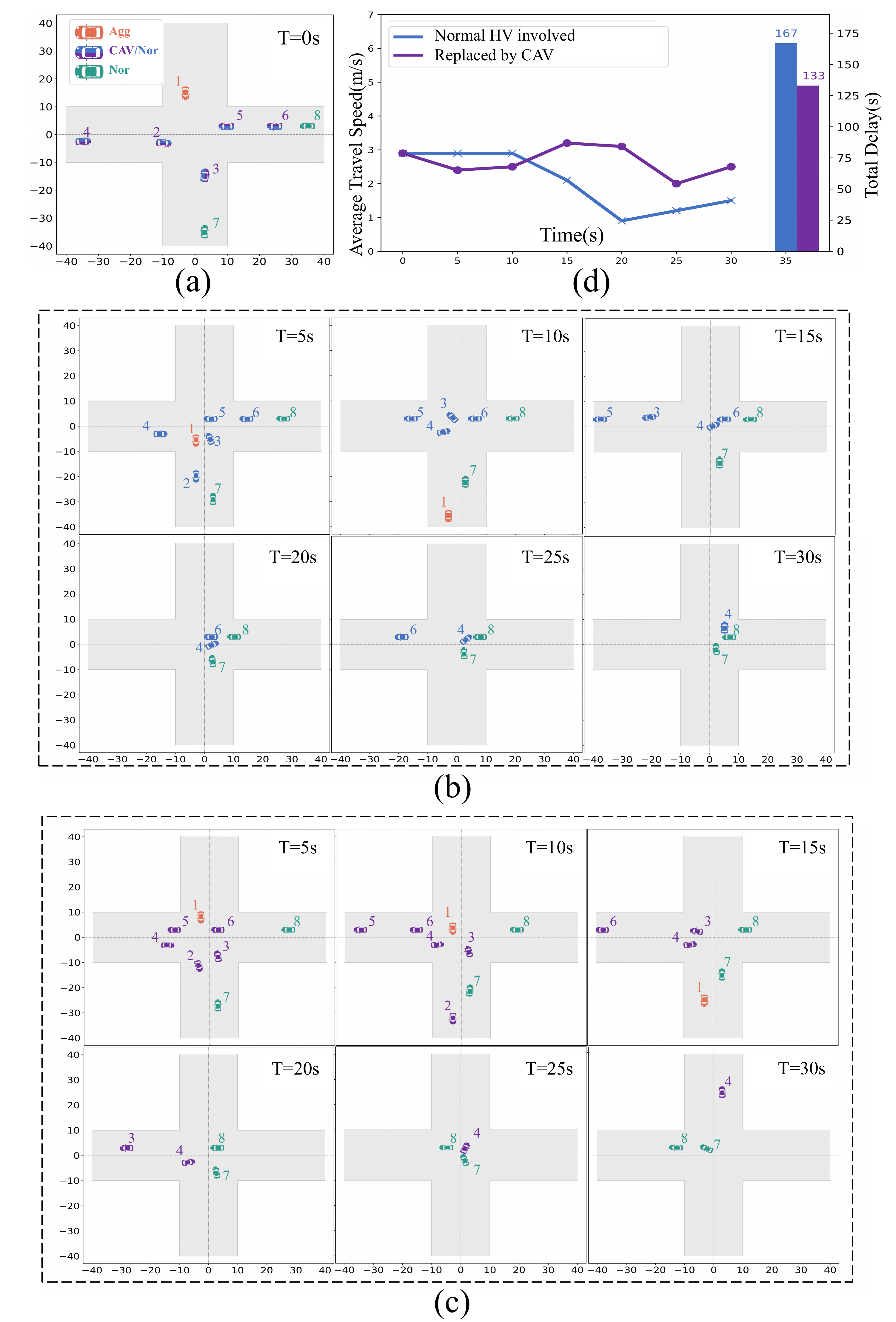}}
  \caption{Interaction case: comparison of normal HV and CAV involved. (a) shows the initial position of the vehicles and the colors of vehicles depict their type; (b) shows the six subsequent steps of the interaction that only HVs participant; (c) shows the corresponding steps after replacing normal HVs with CAVs; (d) summarizes the average travel speed and total delay of this interaction case; numbers next to the vehicles refer to the order in which they enter the simulation environment.}\label{experiment3-8car}
  \end{center}
\end{figure}

From the comparative analysis of Fig.~\ref{experiment1-8car} (d) and Fig.~\ref{experiment3-8car} (d), it can be found that the introduction of heterogeneous HVs has led to a lower average travel speed and higher total delay which brought more challenge to our cooperative driving algorithm framework. Fig.~\ref{experiment3-8car} (b)-(c) shows the process of interaction in detail. In all HV environment, aggressive $V_{1}$ pursued his own benefits and managed to pass as fast as possible. This phenomenon led to a higher average travel speed of the system at the beginning of interaction. While average travel speed plummeted after the aggressive vehicle left the intersection due to its hoggish maneuver that may harm other vehicles in long-range. 

After we replaced normal HVs with CAVs, CAVs were able to actively induce the aggressive vehicle to slow down by showing a willingness to pass first (by comparing the status of Fig.~\ref{experiment3-8car} (b)-(c) at $T=5s$ and $T=10s$). This was attributed to our game-based cooperative method. Their cooperative behaviors were emergent rather than generated by some predefined protocol which gave us the potential to apply to more complex scenarios. Although average travel speed of the whole intersection was slightly lower at first based on Fig.~\ref{experiment3-8car} (d), speed at later simulation duration and total delay are significantly better than without CAVs participation. So far, the effectiveness and robustness of our CAV cooperative driving framework under various environment have been testified.

In conclusion, three simulation cases were conducted for the verification and validation of our cooperative driving framework. The simulation results show that with the designed framework, CAVs are capable of coordinating with heterogeneous vehicles in the high-density, mixed, unsignalized intersection. Besides, as the proportion of CAV increased, the efficiency and safety of the intersection system improved.

% === VI. Conclusion ========================================
% =================================================================================
\section{Conclusion}
With the combination of normalized cooperative game and level-k game, a cooperative driving framework is proposed for CAVs to address the driving conflict in the high-density, mixed, unsignalized intersection.
Differing from reservation-based control methods, the proposed NCL game theoretic approach is capable of cooperating and even actively coordinating with other vehicles. Namely, besides the cooperation between CAVs, CAVs can collaborate with heterogeneous HVs.
Three simulation cases are conducted for verification and validation, including the comparative analysis with different methods, the case study under different ROP and the interaction analysis with heterogeneous HVs. The performance of the intersection system is analyzed through average travel speed, total delay, and PET. Experiment results indicate that the proposed cooperative driving framework is capable of confronting complex, mixed traffic scenarios and therefore nudging system to greater efficiency and safety.

%In addition, driving process consists of long-term personal driving style and short-term behavior, triggered by inherent characteristics and temporary intentions, respectively. Therefore, instead of regarding HVs make decision based on their static personal preference, further investigation should be conducted on the drivers’ time-varying intentions.

% or

% starts a section numbered zero.)the 
%

% ============================================
%\appendices
%Appendix one text goes here %\cite{Roberg2010}.

% iinteractionnt by leaving the argument blank
%\section{}

%\section*{Acknowledgment}

%The authors would like to thank D. Root for the loan of the SWAP. The SWAP that can ONLY be usefull in Boulder...

\ifCLASSOPTIONcaptionsoff
  \newpage
\fi



%\begin{thebibliography}{1}

\footnotesize
\bibliographystyle{IEEEtranN}
\bibliography{IEEEabrv,Bibliography}

% Generated by IEEEtranN.bst, version: 1.14 (2015/08/26)
\begin{thebibliography}{44}
\providecommand{\natexlab}[1]{#1}
\providecommand{\url}[1]{#1}
\csname url@samestyle\endcsname
\providecommand{\newblock}{\relax}
\providecommand{\bibinfo}[2]{#2}
\providecommand{\BIBentrySTDinterwordspacing}{\spaceskip=0pt\relax}
\providecommand{\BIBentryALTinterwordstretchfactor}{4}
\providecommand{\BIBentryALTinterwordspacing}{\spaceskip=\fontdimen2\font plus
\BIBentryALTinterwordstretchfactor\fontdimen3\font minus
  \fontdimen4\font\relax}
\providecommand{\BIBforeignlanguage}[2]{{%
\expandafter\ifx\csname l@#1\endcsname\relax
\typeout{** WARNING: IEEEtranN.bst: No hyphenation pattern has been}%
\typeout{** loaded for the language `#1'. Using the pattern for}%
\typeout{** the default language instead.}%
\else
\language=\csname l@#1\endcsname
\fi
#2}}
\providecommand{\BIBdecl}{\relax}
\BIBdecl
\renewcommand{\BIBentryALTinterwordstretchfactor}{4}

\bibitem[Schwarting et~al.(2019)Schwarting, Pierson, Alonso-Mora, Karaman, and
  Rus]{schwarting2019social}
W.~Schwarting, A.~Pierson, J.~Alonso-Mora, S.~Karaman, and D.~Rus, ``Social
  behavior for autonomous vehicles,'' \emph{Proceedings of the National Academy
  of Sciences}, vol. 116, no.~50, pp. 24\,972--24\,978, 2019.

\bibitem[Camerer and Fehr(2006)]{camerer2006does}
C.~F. Camerer and E.~Fehr, ``When does" economic man" dominate social
  behavior?'' \emph{science}, vol. 311, no. 5757, pp. 47--52, 2006.

\bibitem[Tian et~al.(2018)Tian, Li, Li, Kolmanovsky, Girard, and
  Yildiz]{tian2018adaptive}
R.~Tian, S.~Li, N.~Li, I.~Kolmanovsky, A.~Girard, and Y.~Yildiz, ``Adaptive
  game-theoretic decision making for autonomous vehicle control at
  roundabouts,'' in \emph{2018 IEEE Conference on Decision and Control
  (CDC)}.\hskip 1em plus 0.5em minus 0.4em\relax IEEE, 2018, pp. 321--326.

\bibitem[Xu et~al.(2020)Xu, Zhao, Li, Chen, Kuang, and Zhou]{xu2020nash}
C.~Xu, W.~Zhao, L.~Li, Q.~Chen, D.~Kuang, and J.~Zhou, ``A nash q-learning
  based motion decision algorithm with considering interaction to traffic
  participants,'' \emph{IEEE Transactions on Vehicular Technology}, vol.~69,
  no.~11, pp. 12\,621--12\,634, 2020.

\bibitem[Garz{\'o}n and Spalanzani(2019)]{garzon2019game}
M.~Garz{\'o}n and A.~Spalanzani, ``Game theoretic decision making for
  autonomous vehicles’ merge manoeuvre in high traffic scenarios,'' in
  \emph{2019 IEEE Intelligent Transportation Systems Conference (ITSC)}.\hskip
  1em plus 0.5em minus 0.4em\relax IEEE, 2019, pp. 3448--3453.

\bibitem[Tian et~al.(2020)Tian, Li, Kolmanovsky, Yildiz, and
  Girard]{tian2020game}
R.~Tian, N.~Li, I.~Kolmanovsky, Y.~Yildiz, and A.~R. Girard, ``Game-theoretic
  modeling of traffic in unsignalized intersection network for autonomous
  vehicle control verification and validation,'' \emph{IEEE Transactions on
  Intelligent Transportation Systems}, 2020.

\bibitem[Wang et~al.(2020)Wang, Li, and Zhao]{wang2020performance}
H.~Wang, Y.~Li, and H.~V. Zhao, ``Performance analysis of road intersections
  based on game theory and dynamic level-k model,'' in \emph{2020 IEEE Intl
  Conf on Parallel \& Distributed Processing with Applications, Big Data \&
  Cloud Computing, Sustainable Computing \& Communications, Social Computing \&
  Networking (ISPA/BDCloud/SocialCom/SustainCom)}.\hskip 1em plus 0.5em minus
  0.4em\relax IEEE, 2020, pp. 1112--1119.

\bibitem[Yu et~al.(2018)Yu, Tseng, and Langari]{yu2018human}
H.~Yu, H.~E. Tseng, and R.~Langari, ``A human-like game theory-based controller
  for automatic lane changing,'' \emph{Transportation Research Part C: Emerging
  Technologies}, vol.~88, pp. 140--158, 2018.

\bibitem[Li et~al.(2018)Li, Kolmanovsky, Girard, and Yildiz]{li2018game}
N.~Li, I.~Kolmanovsky, A.~Girard, and Y.~Yildiz, ``Game theoretic modeling of
  vehicle interactions at unsignalized intersections and application to
  autonomous vehicle control,'' in \emph{2018 Annual American Control
  Conference (ACC)}.\hskip 1em plus 0.5em minus 0.4em\relax IEEE, 2018, pp.
  3215--3220.

\bibitem[Karimi and Vahidi(2020)]{karimi2020receding}
S.~Karimi and A.~Vahidi, ``Receding horizon motion planning for automated lane
  change and merge using monte carlo tree search and level-k game theory,'' in
  \emph{2020 American Control Conference (ACC)}.\hskip 1em plus 0.5em minus
  0.4em\relax IEEE, 2020, pp. 1223--1228.

\bibitem[Chu et~al.(2022)Chu, Yu, Yang, and Huang]{chu2022understanding}
P.~Chu, Y.~Yu, J.~Yang, and C.~Huang, ``Understanding the mechanism behind
  young drivers’ distracted driving behaviour based on sor theory,''
  \emph{Journal of Transportation Safety \& Security}, vol.~14, no.~10, pp.
  1655--1673, 2022.

\bibitem[el~abidine Kherroubi et~al.(2021)el~abidine Kherroubi, Aknine, and
  Bacha]{el2021novel}
Z.~el~abidine Kherroubi, S.~Aknine, and R.~Bacha, ``Novel decision-making
  strategy for connected and autonomous vehicles in highway on-ramp merging,''
  \emph{IEEE Transactions on Intelligent Transportation Systems}, 2021.

\bibitem[Yu et~al.(2020)Yu, Lin, Alazab, Tan, and Gu]{yu2020deep}
K.~Yu, L.~Lin, M.~Alazab, L.~Tan, and B.~Gu, ``Deep learning-based traffic
  safety solution for a mixture of autonomous and manual vehicles in a
  5g-enabled intelligent transportation system,'' \emph{IEEE transactions on
  intelligent transportation systems}, vol.~22, no.~7, pp. 4337--4347, 2020.

\bibitem[Yao et~al.(2021)Yao, Zeng, Chen, and He]{yao2021deep}
R.~Yao, W.~Zeng, Y.~Chen, and Z.~He, ``A deep learning framework for modelling
  left-turning vehicle behaviour considering diagonal-crossing motorcycle
  conflicts at mixed-flow intersections,'' \emph{Transportation research part
  C: emerging technologies}, vol. 132, p. 103415, 2021.

\bibitem[Bi et~al.(2012)Bi, Gan, Shang, and Liu]{bi2012queuing}
L.~Bi, G.~Gan, J.~Shang, and Y.~Liu, ``Queuing network modeling of driver
  lateral control with or without a cognitive distraction task,'' \emph{IEEE
  Transactions on Intelligent Transportation Systems}, vol.~13, no.~4, pp.
  1810--1820, 2012.

\bibitem[Huang et~al.(2021)Huang, Wu, and Lv]{huang2021driving}
Z.~Huang, J.~Wu, and C.~Lv, ``Driving behavior modeling using naturalistic
  human driving data with inverse reinforcement learning,'' \emph{IEEE
  Transactions on Intelligent Transportation Systems}, 2021.

\bibitem[Chremos et~al.(2020)Chremos, Beaver, and
  Malikopoulos]{chremos2020game}
I.~V. Chremos, L.~E. Beaver, and A.~A. Malikopoulos, ``A game-theoretic
  analysis of the social impact of connected and automated vehicles,'' in
  \emph{2020 IEEE 23rd International Conference on Intelligent Transportation
  Systems (ITSC)}.\hskip 1em plus 0.5em minus 0.4em\relax IEEE, 2020, pp. 1--6.

\bibitem[Mariani et~al.(2021)Mariani, Cabri, and
  Zambonelli]{mariani2021coordination}
S.~Mariani, G.~Cabri, and F.~Zambonelli, ``Coordination of autonomous vehicles:
  taxonomy and survey,'' \emph{ACM Computing Surveys (CSUR)}, vol.~54, no.~1,
  pp. 1--33, 2021.

\bibitem[Yu et~al.(2019)Yu, Sun, Liu, and Yang]{yu2019managing}
C.~Yu, W.~Sun, H.~X. Liu, and X.~Yang, ``Managing connected and automated
  vehicles at isolated intersections: From reservation-to optimization-based
  methods,'' \emph{Transportation research part B: methodological}, vol. 122,
  pp. 416--435, 2019.

\bibitem[Rios-Torres and Malikopoulos(2016)]{rios2016survey}
J.~Rios-Torres and A.~A. Malikopoulos, ``A survey on the coordination of
  connected and automated vehicles at intersections and merging at highway
  on-ramps,'' \emph{IEEE Transactions on Intelligent Transportation Systems},
  vol.~18, no.~5, pp. 1066--1077, 2016.

\bibitem[Zhou et~al.(2022)Zhou, Ma, Zhao, and Sun]{zhou2022reasoning}
D.~Zhou, Z.~Ma, X.~Zhao, and J.~Sun, ``Reasoning graph: A situation-aware
  framework for cooperating unprotected turns under mixed connected and
  autonomous traffic environments,'' \emph{Transportation Research Part C:
  Emerging Technologies}, vol. 143, p. 103815, 2022.

\bibitem[Xu et~al.(2019)Xu, Zhang, Li, and Li]{xu2019cooperative}
H.~Xu, Y.~Zhang, L.~Li, and W.~Li, ``Cooperative driving at unsignalized
  intersections using tree search,'' \emph{IEEE Transactions on Intelligent
  Transportation Systems}, vol.~21, no.~11, pp. 4563--4571, 2019.

\bibitem[Carlino et~al.(2013)Carlino, Boyles, and Stone]{carlino2013auction}
D.~Carlino, S.~D. Boyles, and P.~Stone, ``Auction-based autonomous intersection
  management,'' in \emph{16th International IEEE Conference on Intelligent
  Transportation Systems (ITSC 2013)}.\hskip 1em plus 0.5em minus 0.4em\relax
  IEEE, 2013, pp. 529--534.

\bibitem[Vu et~al.(2018)Vu, Aknine, and Ramchurn]{vu2018decentralised}
H.~Vu, S.~Aknine, and S.~D. Ramchurn, ``A decentralised approach to
  intersection traffic management.'' in \emph{IJCAI}, 2018, pp. 527--533.

\bibitem[Huang and Nitschke(2020)]{huang2020evolutionary}
C.-L. Huang and G.~Nitschke, ``Evolutionary automation of coordinated
  autonomous vehicles,'' in \emph{2020 IEEE Congress on Evolutionary
  Computation (CEC)}.\hskip 1em plus 0.5em minus 0.4em\relax IEEE, 2020, pp.
  1--7.

\bibitem[Litman(2017)]{litman2017autonomous}
T.~Litman, \emph{Autonomous vehicle implementation predictions}.\hskip 1em plus
  0.5em minus 0.4em\relax Victoria Transport Policy Institute Victoria, BC,
  Canada, 2017.

\bibitem[Wang et~al.(2015)Wang, Hoogendoorn, Daamen, van Arem, and
  Happee]{wang2015game}
M.~Wang, S.~P. Hoogendoorn, W.~Daamen, B.~van Arem, and R.~Happee, ``Game
  theoretic approach for predictive lane-changing and car-following control,''
  \emph{Transportation Research Part C: Emerging Technologies}, vol.~58, pp.
  73--92, 2015.

\bibitem[Sadigh et~al.(2018)Sadigh, Landolfi, Sastry, Seshia, and
  Dragan]{sadigh2018planning}
D.~Sadigh, N.~Landolfi, S.~S. Sastry, S.~A. Seshia, and A.~D. Dragan,
  ``Planning for cars that coordinate with people: leveraging effects on human
  actions for planning and active information gathering over human internal
  state,'' \emph{Autonomous Robots}, vol.~42, no.~7, pp. 1405--1426, 2018.

\bibitem[Vaskov et~al.(2019)Vaskov, Kousik, Larson, Bu, Ward, Worrall,
  Johnson-Roberson, and Vasudevan]{vaskov2019towards}
S.~Vaskov, S.~Kousik, H.~Larson, F.~Bu, J.~Ward, S.~Worrall,
  M.~Johnson-Roberson, and R.~Vasudevan, ``Towards provably not-at-fault
  control of autonomous robots in arbitrary dynamic environments,'' \emph{arXiv
  preprint arXiv:1902.02851}, 2019.

\bibitem[Liu et~al.(2022)Liu, Wan, Lewis, Nageshrao, and Filev]{liu2022three}
M.~Liu, Y.~Wan, F.~L. Lewis, S.~Nageshrao, and D.~Filev, ``A three-level
  game-theoretic decision-making framework for autonomous vehicles,''
  \emph{IEEE Transactions on Intelligent Transportation Systems}, 2022.

\bibitem[Ni et~al.(2016)Ni, Wang, Sun, and Li]{ni2016evaluation}
Y.~Ni, M.~Wang, J.~Sun, and K.~Li, ``Evaluation of pedestrian safety at
  intersections: A theoretical framework based on pedestrian-vehicle
  interaction patterns,'' \emph{Accident Analysis \& Prevention}, vol.~96, pp.
  118--129, 2016.

\bibitem[Chen et~al.(2019)Chen, Chen, and Chen]{chen2019driving}
K.-T. Chen, H.-Y.~W. Chen, and H.-Y.~W. Chen, ``Driving style clustering using
  naturalistic driving data,'' \emph{Transportation research record}, vol.
  2673, no.~6, pp. 176--188, 2019.

\bibitem[Huang et~al.(2018)Huang, Sun, and Zhang]{huang2018effects}
Y.~Huang, D.~J. Sun, and L.-H. Zhang, ``Effects of congestion on drivers’
  speed choice: Assessing the mediating role of state aggressiveness based on
  taxi floating car data,'' \emph{Accident Analysis \& Prevention}, vol. 117,
  pp. 318--327, 2018.

\bibitem[Rahmati et~al.(2021)Rahmati, Hosseini, and
  Talebpour]{rahmati2021helping}
Y.~Rahmati, M.~K. Hosseini, and A.~Talebpour, ``Helping automated vehicles with
  left-turn maneuvers: a game theory-based decision framework for conflicting
  maneuvers at intersections,'' \emph{IEEE Transactions on Intelligent
  Transportation Systems}, 2021.

\bibitem[Cheng et~al.(2019)Cheng, Yao, Zhang, Li, and Guo]{cheng2019vehicle}
C.~Cheng, D.~Yao, Y.~Zhang, J.~Li, and Y.~Guo, ``A vehicle passing model in
  non-signalized intersections based on non-cooperative game theory,'' in
  \emph{2019 IEEE Intelligent Transportation Systems Conference (ITSC)}.\hskip
  1em plus 0.5em minus 0.4em\relax IEEE, 2019, pp. 2286--2291.

\bibitem[Ziebart et~al.(2008)Ziebart, Maas, Bagnell, Dey,
  et~al.]{ziebart2008maximum}
B.~D. Ziebart, A.~L. Maas, J.~A. Bagnell, A.~K. Dey \emph{et~al.}, ``Maximum
  entropy inverse reinforcement learning.'' in \emph{Aaai}, vol.~8.\hskip 1em
  plus 0.5em minus 0.4em\relax Chicago, IL, USA, 2008, pp. 1433--1438.

\bibitem[Wang et~al.(2022)Wang, Zhang, and Peng]{wang2022comprehensive}
X.~Wang, S.~Zhang, and H.~Peng, ``Comprehensive safety evaluation of highly
  automated vehicles at the roundabout scenario,'' \emph{IEEE Transactions on
  Intelligent Transportation Systems}, 2022.

\bibitem[Dresner and Stone(2008)]{dresner2008multiagent}
K.~Dresner and P.~Stone, ``A multiagent approach to autonomous intersection
  management,'' \emph{Journal of artificial intelligence research}, vol.~31,
  pp. 591--656, 2008.

\bibitem[Yang et~al.(2018)Yang, Huang, Wang, Pei, and Yao]{yang2018cooperative}
Z.~Yang, H.~Huang, G.~Wang, X.~Pei, and D.-y. Yao, ``Cooperative driving model
  for non-signalized intersections with cooperative games,'' \emph{Journal of
  Central South University}, vol.~25, no.~9, pp. 2164--2181, 2018.

\bibitem[Xing et~al.(2022)Xing, Su, Xu, Zhang, and Luan]{xing2022secure}
R.~Xing, Z.~Su, Q.~Xu, N.~Zhang, and T.~H. Luan, ``Secure content delivery for
  connected and autonomous trucks: A coalition formation game approach,''
  \emph{IEEE Transactions on Intelligent Transportation Systems}, 2022.

\bibitem[Guo et~al.(2020)Guo, Cheng, and Liu]{guo2020merging}
J.~Guo, S.~Cheng, and Y.~Liu, ``Merging and diverging impact on mixed traffic
  of regular and autonomous vehicles,'' \emph{IEEE Transactions on Intelligent
  Transportation Systems}, vol.~22, no.~3, pp. 1639--1649, 2020.

\bibitem[Greenshields et~al.(1935)Greenshields, Bibbins, Channing, and
  Miller]{greenshields1935study}
B.~Greenshields, J.~Bibbins, W.~Channing, and H.~Miller, ``A study of traffic
  capacity,'' in \emph{Highway research board proceedings}, vol. 1935.\hskip
  1em plus 0.5em minus 0.4em\relax National Research Council (USA), Highway
  Research Board, 1935.

\bibitem[Allen et~al.(1978)Allen, Shin, and Cooper]{allen1978analysis}
B.~L. Allen, B.~T. Shin, and P.~J. Cooper, ``Analysis of traffic conflicts and
  collisions,'' Tech. Rep., 1978.

\bibitem[Qi et~al.(2020)Qi, Wang, Shen, and Wu]{qi2020modified}
W.~Qi, W.~Wang, B.~Shen, and J.~Wu, ``A modified post encroachment time model
  of urban road merging area based on lane-change characteristics,'' \emph{IEEE
  Access}, vol.~8, pp. 72\,835--72\,846, 2020.

\end{thebibliography}

\vfill


\end{document}